# Single-molecule fluorescence multiplexing by multi-parameter spectroscopic detection of nanostructured FRET labels

**Jiachong Chu[1*], Ayesha Ejaz[2*], Kyle M. Lin[3,4], Madeline R. Joseph[1], Aria E. Coraor[1], D. Allan Drummond[5,6,7], and Allison H. Squires[1,7⌂]**

[1]Pritzker School of Molecular Engineering, University of Chicago, IL, USA
[2]Department of Chemistry, University of Chicago, IL, USA
[3]Graduate Program in Biophysical Sciences, University of Chicago, Chicago, IL, USA
[4]Interdisicplinary Scientist Training Program, Pritzker School of Medicine, University of Chicago, Chicago, IL, USA
[5]Department of Biochemistry and Molecular Biophysics, University of Chicago, Chicago, IL, USA
[6]Department of Medicine, Section of Genetic Medicine, University of Chicago, Chicago, IL, USA
[7]Institute for Biophysical Dynamics, University of Chicago, Chicago, IL, USA
*Authors contributed equally to this work
⌂e-mail correspondence: asquires@uchicago.edu

## Abstract

Multiplexed, real-time fluorescence detection at the single-molecule level is highly desirable to reveal the stoichiometry, dynamics, and interactions of individual molecular species within complex systems. However, traditionally fluorescence sensing is limited to detection of 3-4 labels at a time, due to low signal-to-noise, high spectral overlap between labels, and the need to avoid dissimilar dye chemistries. To surmount these barriers, we have engineered a palette of several dozen fluorescent labels, called FRETfluors, for spectroscopic multiplexing at the single-molecule level. Each FRETfluor is a compact nanostructure formed from the same three chemical building blocks (DNA, Cy3, and Cy5). The composition and dye-dye geometries create a characteristic Förster Resonance Energy Transfer (FRET) efficiency for each construct. In addition, we varied the local DNA sequence and attachment chemistry to alter the Cy3 and Cy5 emission properties and thereby shift the emission signatures of an entire series of FRET constructs to new sectors of the multi-parameter detection space. Unique spectroscopic emission of each FRETfluor is therefore conferred by a combination of FRET and this site-specific tuning of individual fluorophore photophysics. We show simultaneous single-molecule identification of a set of 27 FRETfluors in a sample mixture using a subset of constructs statistically selected to minimize classification errors, measured using an Anti-Brownian ELectrokinetic (ABEL) trap which provides precise multi-parameter spectroscopic measurements. The ABEL trap also reveals transport properties of a trapped particle, which enables discrimination between FRETfluors attached to a target and unbound FRETfluors, eliminating the need for washes or removal of excess label by purification. Finally, we demonstrate detection of both simple and complex mixtures of mRNA, dsDNA, and proteins, providing proof-of-concept for applications to amplification-free sensing of low-abundance targets in highly heterogeneous samples. Although usually considered an undesirable complication of fluorescence, here the inherent sensitivity of fluorophores to the local physicochemical environment provides a new design axis that is nearly orthogonal to changing the geometry of a FRET construct. As a result, the number of distinguishable FRET-based labels can be combinatorically expanded while maintaining chemical compatibility, opening up new possibilities for spectroscopic multiplexing at the single-molecule level using a minimal set of chemical components.

## Introduction

Multiplexed measurements provide critical insights into the molecular compositions and interactions that govern complex nanoscale systems. Fluorescent labels offer highly sensitive and specific readout of molecular identity, enabling information-rich imaging and assays with a rainbow of colors for microscale objects.[1,2] At the single-molecule level, color ratio-based multiplexing has been demonstrated by multiple groups for up to ~10 labels exhibiting unique color combinations,[3] and up to 25 labels with a complex scheme of four-laser alternating excitation and four dyes.[4] However, low signal-to-noise ratios and overlapping spectra generally restrict single-molecule fluorescence multiplexing on the vast majority of microscopes to at most 3-4 colors[5]. Currently, options for color ratio-based multiplexing remain limited at the single-molecule level.

To overcome these limitations, technologies for single-molecule fluorescence multiplexing must utilize additional measurement dimensions to separate signals. Chemical fixation and surface immobilization of samples enable elegant multiplexing strategies for up to thousands of labels via barcoding approaches where unique combinations of molecular interactions are detected through multiple rounds of readout for each molecule, providing patterning in temporal[6–8], spatial[9–13], or kinetic[14–16] dimensions. For living or dynamic samples, however, multiplexing must be encoded in additional spectroscopic dimensions for each fluorescent label without the aid of temporal or spatial patterning, since repeated measurements of the same molecule cannot be guaranteed.[17] Spectroscopic properties including fluorescence brightness and quantum yield, fluorescence lifetime, anisotropy, and emission spectrum are routinely accessible with single-molecule sensing methods.[18–22] By labeling samples with fluorophores that possess different photophysical properties, spectroscopically multiplexed imaging at or approaching the single molecule level has been demonstrated using up to nine labels,[23] and other sensing modalities have achieved up to six labels in complex sample mixtures with different fluorophores.[3,24] But although use of chemically diverse fluorophores offers a potentially broad spectroscopic palette, the number of labels that can be concurrently identified is ultimately constrained by dissimilar chemical compatibility and labeling performance across different chemical structures.

One well-established means to generate a variety of spectroscopic signals using a limited number of compounds is Förster Resonance Energy Transfer (FRET) between fluorophores positioned on DNA nanostructures.[25–27] The pairwise rate of energy transfer between an excited donor and a potential acceptor is influenced by the photophysical properties of both fluorophores and by their relative geometry, including spatial separation and orientation.[5,28] DNA nanotechnology is a powerful tool for engineering FRET networks because it can be used to create molecular scaffolds that position and orient covalently linked fluorophores with sub-nanometer precision,[9,29] governing the flow of energy through the structure and thereby dictating the observed spectroscopic properties.[30–32] (Note that many barcoding-based single-molecule fluorescence multiplexing approaches also employ DNA-fluorophore constructs to achieve spatial or interaction patterning.) For a single FRET pair (one donor, one acceptor) on a simple DNA scaffold, spectroscopic multiplexing using several DNA constructs[33] and up to as many as 15 constructs,[34] has been demonstrated at the single-molecule level. In theory, further multiplexing could be achieved using more complex DNA-FRET constructs, or constructs with additional donors, acceptors, or fluorophore types. However, practical limitations of DNA-FRET constructs for multiplexing must be considered, including the typical spatial extent of FRET interactions (~10 nm) and tradeoffs among scaffold complexity, spectroscopic uniqueness, and error-free assembly and readout. Moreover, emission from a FRET network reflects tight coupling among spectroscopic properties due to underlying physical processes, so commonly measured variables such as donor lifetime and brightness, FRET efficiency, emission spectrum, and acceptor brightness show correlated or anticorrelated variation across different constructs.[5]

Dyes attached to DNA are sensitive to changes in local base sequence and attachment chemistries, both of which influence a dye's physicochemical environment by changing structural flexibility and conformations as well as solvent accessibility and interactions with the DNA scaffold.[35] Cyanine dyes on DNA are a particularly well-studied class of constructs in this context,[36] and exhibit changing lifetimes and quantum yields,[37–40] orientation and base stacking,[27,41–43] system-bath coupling,[44] and torsion and isomerization.[45–47] These photophysical changes directly impact energy transfer within DNA-FRET nanostructures.[37,44]

Here we demonstrate that photophysical modifications can be engineered into DNA-FRET constructs to facilitate spectroscopic multiplexing at the single-molecule level. We have designed a set of labels, called "FRETfluors," which produce unique spectroscopic signatures from three simple building blocks (DNA, Cy3, and Cy5). Energy transfer between the donor (Cy3) and acceptor (Cy5), in combination with the effects of local DNA sequence and attachment chemistry, tunes the spectroscopic emission of each FRETfluor across multiple measurable parameters including color, brightness, and fluorescence lifetime. Readout in an Anti-Brownian ELectrokinetic (ABEL) trap[48–52] provides high-precision multiparameter measurements for FRETfluors in solution at sub-picomolar concentrations. We show sequence-specific labeling of ssDNA, dsDNA, mRNA, and proteins with FRETfluors, and show that it is not necessary to wash or purify out excess labels that have not found a target, because the ABEL trap reliably discriminates between target-bound and free FRETfluors.[34,53] Pairwise comparison of the characteristic emission parameters observed for each FRETfluor allows statistically optimal subsets to be selected for multiplexing, demonstrated here with a set of 27 FRETfluor labels in a single mixture. Finally, we show proof-of-concept for applications to detection of low-abundance biomolecular targets in highly heterogeneous biomolecular samples, illustrating that combining FRET with tunable fluorophore photophysics provides new opportunities for signal multiplexing from a minimal set of chemical components.

## Design of FRETfluor labels

We sought to create a collection of fluorescent labels with unique spectroscopic signals, high chemical homogeneity, and minimal structural complexity. FRETfluors use a minimal set of biomolecular building blocks: DNA oligomers functionalized with either Cy3 or Cy5 dye (see Supplementary Table S1 for sequences), with a total molecular weight of about 30 kDa, similar to a fluorescent protein but with a higher aspect ratio. We first created a series of constructs incorporating non-sulfonated Cy3 and Cy5 into the DNA backbone as shown in Fig. 1a. Cy3 is positioned in the "A" strand (cyan), with $N$ base pairs (here: 9 bp) separating it from the Cy5 positioned in the "B" strand (blue). Hereafter these constructs are referred to as "AB$N$" (here: AB9). We chose phosphoramidite incorporation of dyes into the DNA backbone to limit dipole rotational mobility[38] and to improve photostability.[54] Other AB$N$ FRETfluors have Cy3 in the same location, with Cy5 placed closer or farther for $6 \leq N \leq 20$ (spacing between 2 and 7 nm). A "bridge" strand (green) hybridizes to the 3' end of the A strand, with an overhang whose sequence can be tuned to match a target nucleic acid, enabling sequence-specific labeling of the target. The single-exponential fitted lifetime and background-subtracted brightness of the Cy3 donor in an AB$N$ complex lacking Cy5 (AB0 construct) were measured to be $\tau_{AB0}$ = 1.6 ± 0.03 ns, green brightness 0.31 ± 0.012 counts ms$^{-1}$ µW$^{-1}$.

Although FRET can be used to tune the emission profile across a set of fluorophores, the resulting spectroscopic variables (brightness, lifetime, emission spectrum or FRET efficiency) are highly coupled. For the AB*N* series of constructs, we expected to measure spectroscopic states on a smooth manifold in this parameter space, as has been previously reported by others.[27,34] To access different sectors of the spectroscopic detection space, this manifold must be shifted by altering the photophysical properties of one or both fluorophores. By changing only sequence and attachment chemistry of the Cy3 donor or by including an additional Cy3, we created three additional label types with different photophysical properties from AB*N*, shown in Fig. 1b with key differences circled (red dotted). The "skip" oligos, $B_{sk}$, are modified B strands that lack the unpaired bases opposing Cy3 and Cy5 (compare with Fig. 1a, orange), lowering the lifetime and quantum efficiency of the donor Cy3 as compared to the AB*N* series ($\tau_{ABsk0}$ = 1.25 ns ± 0.03, green brightness 0.26 ± 0.007 counts ms$^{-1}$ µW$^{-1}$). The "cap" oligos, $A_c$, are modified A oligos that carry an additional single-tethered Cy3 at the 5' end, increasing total brightness and lowering net Cy3 lifetime ($\tau_{AcB0}$ = 1.07 ns ± 0.03, green brightness 0.40 ± 0.011 counts ms$^{-1}$ µW$^{-1}$). The "internal" oligos, $B_{in}$, incorporate an additional Cy3 between the 3' end of the bridge strand and the 5' end of the B strand, where it acts as an additional donor to increase brightness with near-normal Cy3 lifetime ($\tau_{ABin0}$ = 1.51 ns ± 0.03, green brightness 0.56 ± 0.015 counts ms$^{-1}$ µW$^{-1}$). (See Supplementary Note S1, Supplementary Table S2, and Supplementary Fig. S1 for additional details, data, and discussion of donor-only photophysics.)

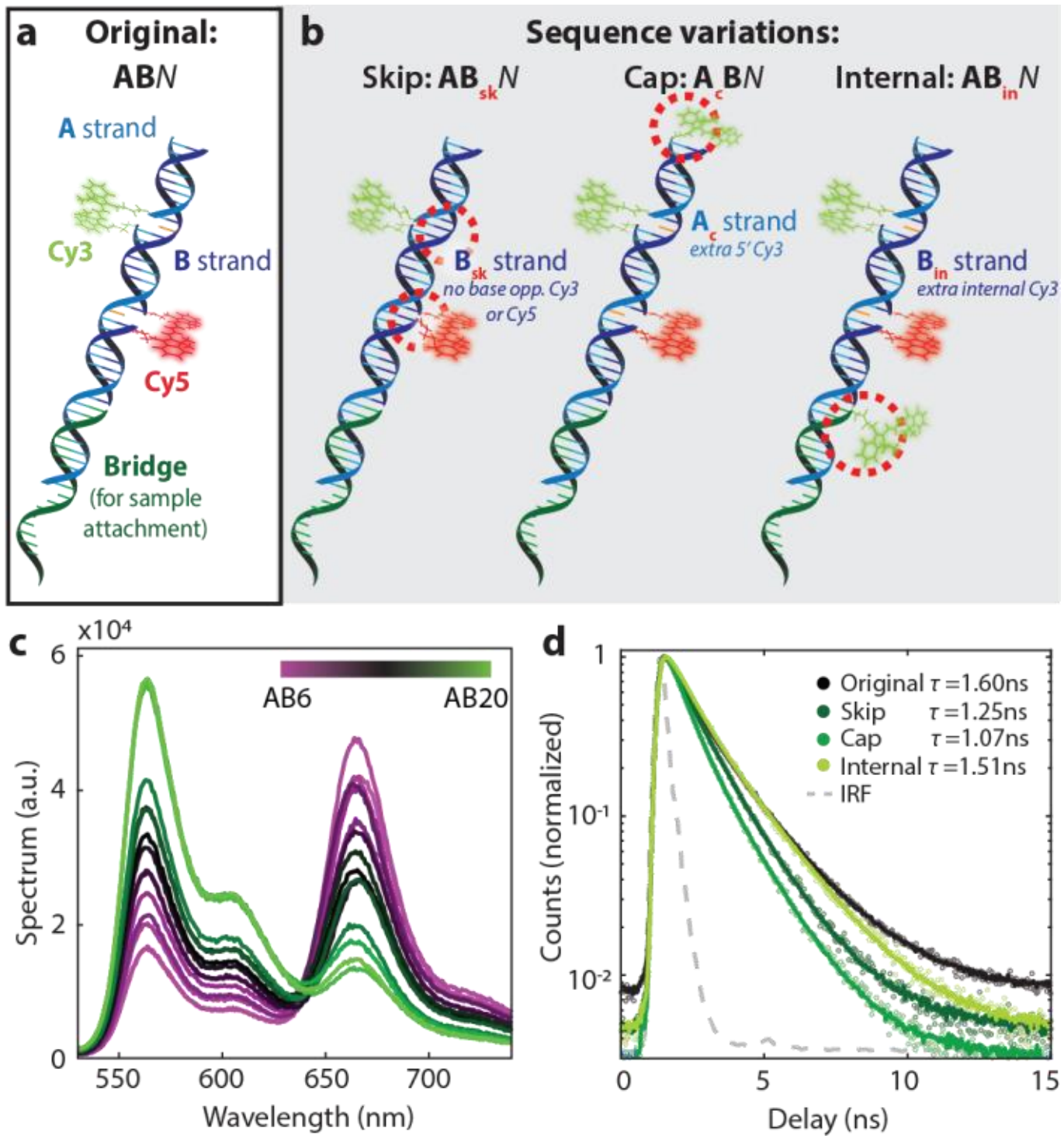

**Figure 1.** FRETfluor concept and design. a) FRETfluor design for AB*N* constructs, with a bridge for sequence-specific labeling. b) Oligo sequence and design variations $AB_{sk}$, $A_cB$, and $AB_{in}$, used to create additional unique spectroscopic signatures. Key changes for each construct are highlighted with a dotted red circle. c) Bulk emission spectra of AB*N* constructs demonstrate that FRET tunes the emission as expected. d) Fluorescence lifetime measurements from single molecules show that Cy3's lifetime depends upon the local DNA sequence (1-exp fits shown; IRF in gray dotted).

Bulk emission spectra of 15 different AB*N* constructs are plotted in Fig. 1c, illustrating the expected increase in Cy5 emission with decreasing *N*. In Fig. 1d, single-exponent fits to the measured lifetime decays of Cy3 for each type of construct are shown as described above, illustrating the effect of local sequence and attachment chemistry on donor lifetime. In total, 41 FRETfluor constructs were synthesized: 15 of AB*N*, 8 of $AB_{sk}N$, 9 of $A_cBN$, and 9 of $AB_{in}N$. Most, but not all, constructs were uniquely identifiable at the single molecule level; *vide infra*.

## Detection of FRETfluor labels in the ABEL trap

To spectroscopically characterize the single-molecule emission of each FRETfluor, we employed a custom-built ABEL trap (Fig. 2a). The ABEL trap applies closed-loop feedback voltages to electrophoretically counteract the effects of Brownian motion in a solution-phase environment. Originally developed by Cohen and Moerner [48,55] to overcome common technical challenges in single-molecule measurements, the ABEL trap maintains the position of a single molecule conjugate to a point detector for extended time, allowing for an isotropic view under constant illumination without the need for tethers or surfaces. Additional details of ABEL trap operation are available in multiple reviews.[56–58] Critically, ABEL traps enable spectroscopic characterization of single molecules across multiple parameters, including brightness, fluorescence lifetime, anisotropy, and emission spectrum.[49,59,60,51,52,61,22] With the recorded applied voltages, trapped particles' observed locations over time can be used to estimate hydrodynamic properties of single particles, including diffusion coefficient and electrophoretic mobility, allowing us to monitor FRETfluor size and charge during trapping.[53]

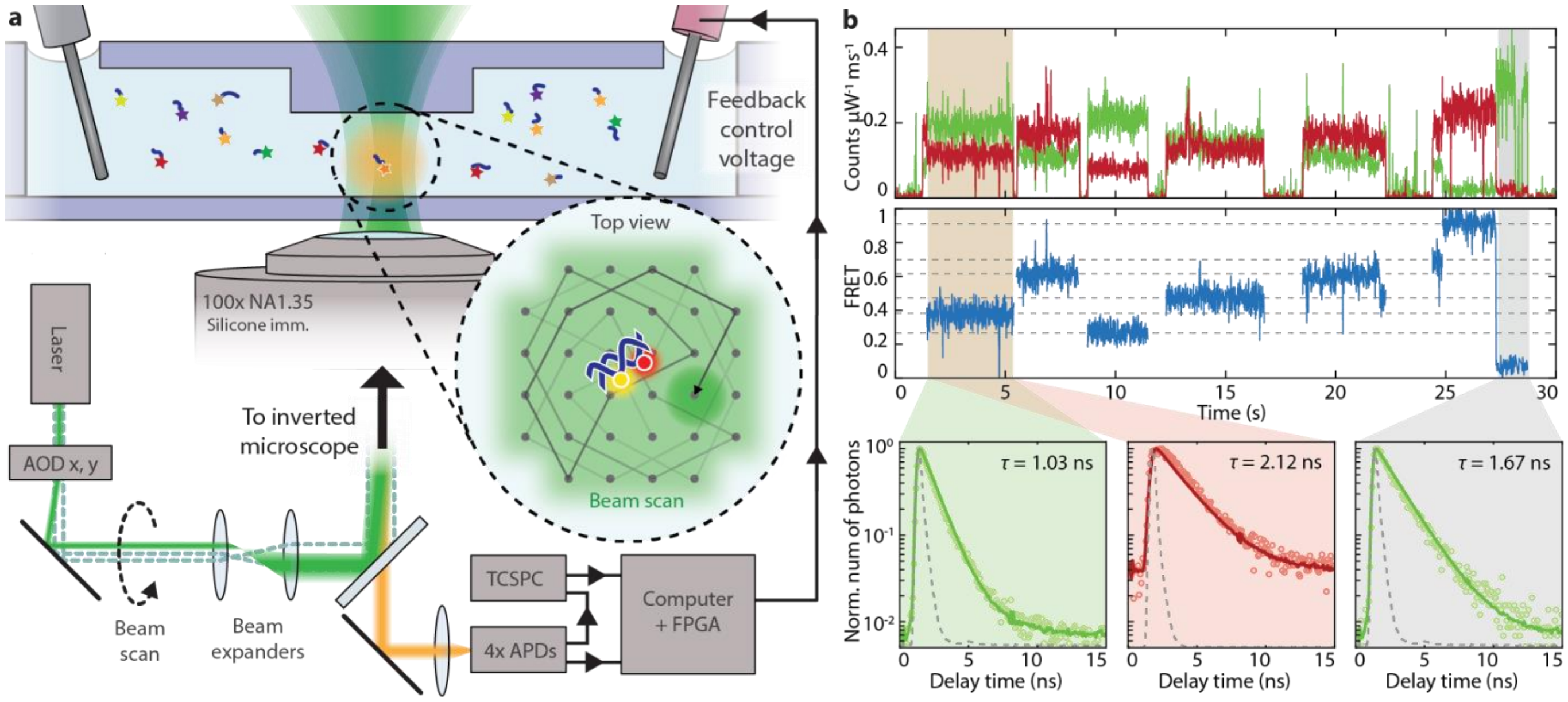

**Figure 2.** ABEL trap-based detection of FRETfluors in a complex sample. a) Schematic of ABEL trap detection: FRETfluors (blue DNA, colored stars), are detected in a microfluidic cell atop an inverted microscope. 532 nm laser excitation is scanned across the field of view using *x* and *y* acousto-optic deflectors (AODs). A FRETfluor in the trapping region fluoresces when it is co-localized with the scanned laser position, enabling closed-loop feedback control over its position via electrodes that apply *x* and *y* voltages to electrophoretically move the particle back to trap center. Spectroscopic data is simultaneously acquired. b) Raw ABEL trap data showing signals from 7 different FRETfluors over 30 seconds. *Top:* background-subtracted brightness in red and green channels is observed during trapping. *Middle:* FRET efficiency calculated from red and green brightness. Gray dotted lines indicate expected FRET values for each class of FRETfluor. *Bottom:* Fluorescence lifetime decays for green and red channels during first trapping event (brown → green and red backgrounds) vs. when the acceptor is blinking or photobleached (gray background). The IRF is shown in dotted gray.

In our ABEL trap setup (Fig. 2a), fluorescence from a trapped molecule is collected in red and green emission channels with concurrent polarization and lifetime information. During each individual trapping event shown in Fig. 2b, the FRETfluor identity can be determined by a combination of the observed parameters (expected FRET values for each FRETfluor shown are calculated from clustering analysis; *vide infra*). From the lifetime fits, we confirm that the donor lifetime is substantially shortened when energy is being transferred to Cy5 (Fig. 1b, bottom left). At a later time (highlighted in gray), the acceptor Cy5 has either blinked off temporarily, or photobleached, so that we see only the donor Cy3 signal. It is clear that each event exhibits different green and red brightness levels, and different FRET values, (expected FRET values for each type of FRETfluor included for reference, see dotted gray lines). Additional raw ABEL trap data with FRETfluor identity annotated for each trapping event are shown in Supplementary Fig. S2 and discussed in Supplementary Note S2.

The typical trapping rate for FRETfluors in our ABEL trap setup is ~0.1 molecules/s·pM. At this trapping rate, with a measurement time of just ~15 min we can detect FRETfluor-labeled samples at ultra-low concentrations down to tens of femtomolar (see Supplementary Fig. S3 and Supplementary Note S3 for details). The ABEL trap's high sensitivity to extremely low concentrations of FRETfluors is a major advantage of our approach.

## Wash-free labeling of specific biomolecular targets

Diverse attachment chemistries, such as small ligands, nucleic acids, or peptides, can be conjugated to a FRETfluor DNA oligo. The specificity of FRETfluors to common biomolecular targets can be tuned as for other fluorescent labeling strategies, e.g.: antibodies, specific chemical linkers, or sequence complementarity. Here, we demonstrate sequence-specific labeling of nucleic acids including ssDNA, mRNA, and dsDNA, as well as site-specific labeling of proteins.

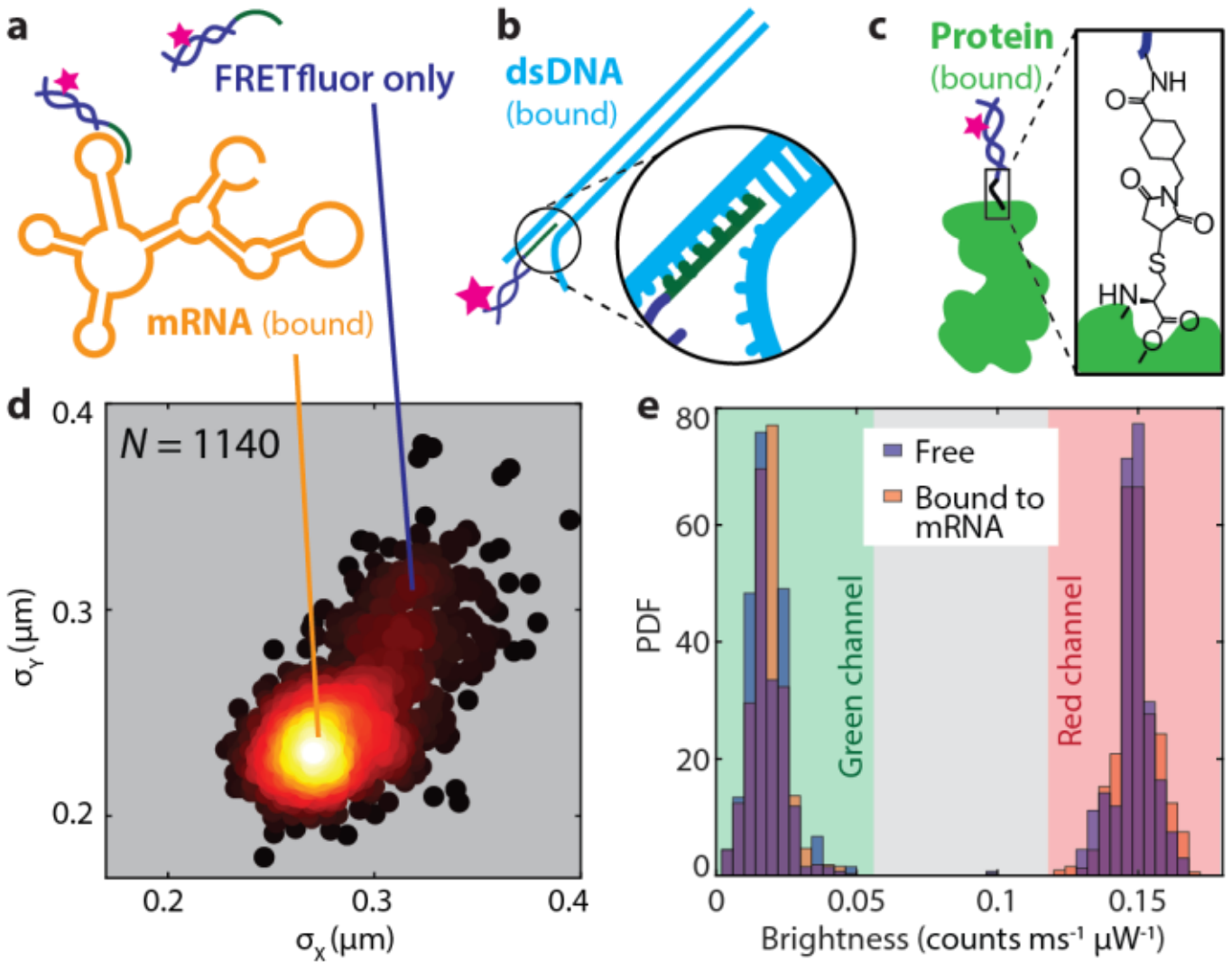

**Figure 3.** Sequence-specific labeling of mRNA, dsDNA, and proteins by FRETfluors. a) Illustration of a free FRETfluor tag (upper) and FRETfluor targeting to a specific mRNA (lower). b) Illustration of a FRETfluor tag targeted to dsDNA. c) Illustration of FRETfluor site-specifically labeling a protein via a maleimide-NSH ester linker. d) Scatter plot of standard deviation of position in *x* and *y* for trapped molecules shows two populations. Points are colored according to the local relative scatter plot density. e) Normalized histograms of green (left) and red (right) signals are nearly identical between free (blue) and mRNA-bound (orange) FRETfluors.

We targeted FRETfluors to specific nucleic acids via sequence complementarity between a single-stranded bridge strand (Fig. 1a and 1b, Fig. 3a and 3b, green) on the FRETfluor and the DNA or RNA target. We verified the FRETfluors' target specificity for nucleic acids by labeling ssDNA oligomers in a bulk electrophoretic mobility shift assay (EMSA; shown in Supplementary Fig. S4). No binding was observed for a bridge sequence lacking complementarity to the ssDNA target ($BR_{off\text{-}target}$), while nearly 100% binding was observed with the correct bridge sequence ($BR_{target}$). We next tested binding to three different mRNAs by designing bridge sequences complementary to regions predicted with high confidence to be part of a loop in the secondary structure of: enhanced Green Fluorescent Protein mRNA (EGFP; 996 nt), firefly luciferase mRNA (FLuc; 1929 nt), and ovalbumin mRNA (OVA; 1438 nt).[62] Bulk EMSAs showed that the off-target bridge sequence could not bind mRNA, while the on-target sequence was correctly hybridized for each mRNA tested (see Supplementary Note S4 and Supplementary Fig. S5). We similarly confirmed by EMSA that a FRETfluor with a correct targeting bridge sequence could invade and bind near the end of a dsDNA RT-PCR product (see Supplementary Fig. S6). To site-specifically label proteins with FRETfluors, we utilized a bifunctional linker containing both an NHS ester group and a maleimide group, which we reacted with a primary amine on a FRETfluor and a sulfhydryl group on the target protein as illustrated in Fig. 3c. We verified covalent labeling by EMSA for two target proteins, poly-A binding protein (Pab1) and a Class A J-domain protein (Ydj1) from *S. cerevisiae*, mutated to have a single accessible cysteine (see Supplementary Fig. S7).

For labeling applications, it is essential to separate or distinguish free labels from target-bound labels. Usually, fluorescence labeling protocols require substantial washing or sample purification, since a correctly bound label cannot be distinguished from an unbound or free tag on the basis of brightness or other spectroscopic properties alone. With the ABEL trap, on-target labeling of high molecular-weight samples such as mRNA can be readily confirmed via the measured transport properties of the trapped object. Objects with a larger hydrodynamic radius, such as a labelled target, have more charge and will in principle diffuse more slowly than free labels due to larger size, leading to tighter confinement around the trap's center. The scatter plot in Fig. 3d shows the positional deviation (calculated for 1000-photon bins) from trap center in each direction, $\sigma_x$ and $\sigma_y$, for many trapping events for FLuc mRNA labeled with AB6 FRETfluor. FRETfluors that have correctly attached to a target mRNA (orange) form a more confined population, with smaller $\sigma_x$ and $\sigma_y$ compared to the large population of free AB6 labels (blue). The spectroscopic signals observed for labeled targets and free FRETfluors are nearly identical, as shown in Fig. 3e for the labeled FLuc mRNA, so FRETfluors can be reliably identified whether free or bound to a target. The change in trapping confinement due to increased size, in combination with minimal or no change to the FRETfluor signal, enables wash-free detection of FRETfluor-labeled biomolecules in the ABEL trap.

## Multi-parameter characterization of FRETfluor emission

To learn which measured variables captured the most useful information for FRETfluor identification, we collected and aggregated photon-by-photon brightness and lifetime data in four channels (red/green, parallel/perpendicular polarization) from many trapping events. Maximum-likelihood changepoints were identified as described in *Methods: Analysis* to parse the data into discrete levels of constant red and green emission, and only levels with durations > 150 ms were considered for further analysis. Levels passing this filter contain an average of more than 5000 photons (see Supplementary Fig. S8). As seen in Fig. 4 for many different constructs, each FRETfluor exhibits tightly-clustered, self-consistent spectroscopic emission levels. We found that analyzing donor lifetime, FRET efficiency, and red and green channel brightnesses as orthogonal dimensions led to optimal separation and identification of FRETfluor populations.

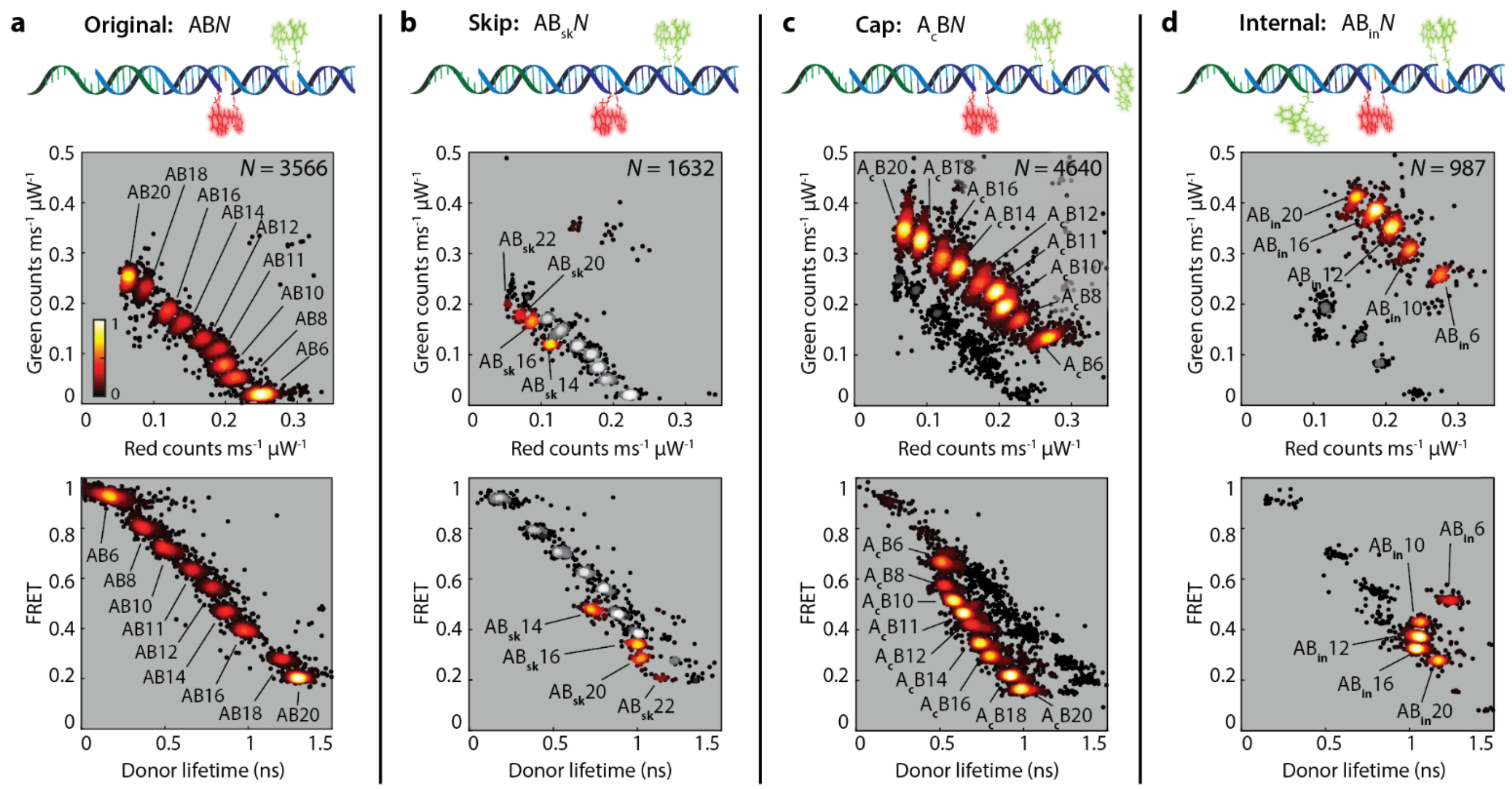

**Figure 4.** Tuning Cy3 photophysics shifts spectroscopic properties of FRETfluor labels. Red-green and lifetime-FRET projections of level-by-level data from trapped FRETfluor constructs show clusters in different regions of the measured multi-parameter space. a) A set of 9 AB*N* constructs show distinct clusters in both red-green (top) and lifetime-FRET projections (bottom). b) Data for four AB$_{sk}$*N* constructs was taken along with nine AB*N* constructs (grayscale) to verify shifted cluster locations. c) Data for nine A$_c$B*N* constructs and d) five AB$_{in}$*N* constructs similarly show distinct clusters that are distinguishable from the original AB*N* construct locations. Throughout: The black-red-yellow heatmap shows relative scatter plot density from low (black) to high (yellow).

We tested FRETfluors individually and in various combinations to determine the characteristic emission properties and cluster widths for each construct. Fitted average properties and standard deviations are listed for all 41 constructs in Supplementary Table S3. Fig. 4a shows two different 2-D projections of data from a mixture of nine different AB*N* labels. The top projection shows red vs. green brightness ("red-green projection"), and as expected for a series of FRET constructs, these values are approximately inversely correlated. The bottom projection shows the values for single-exponential lifetime fits for the Cy3 donor in the parallel channel vs. FRET efficiency ("lifetime-FRET" projection), which similarly shows an inverse correlation. Nine distinct clusters are clearly evident in each of these projections. Clusters are labeled with the corresponding construct name, as verified by additional experiments using different combinations of constructs, as well as single construct experiments (not shown). The remaining six AB*N* constructs were determined to statistically overlap with one or more of the constructs shown here beyond a 2.5% threshold for probability of misclassification of either FRETfluor label (see *Selection of FRETfluor sets for robust classification*), and therefore are not included here (AB7, AB9, AB13, AB15, AB17, AB19). Data for these clusters are shown in Supplementary Fig. S9.

We characterized all other FRETfluor constructs to learn how changes to the donor photophysics shifted the manifold of FRET states observed in the data. Fig. 4b-d show the clusters for combinations from each of the other construct types. Four of eight AB$_{sk}$*N* constructs showed unique signals relative to the AB*N* constructs and one another (Fig. 4b). A subset of the nine AB*N* constructs from Fig. 4a was included in this mixture to verify that the $AB_{sk}N$ constructs were indeed distinguishable in a mixture (shown in grayscale). Nine $A_cBN$ constructs and five $AB_{in}N$ constructs were also readily distinguishable relative to the original AB*N* constructs (Fig. 4c and d, respectively). Data for all remaining constructs, which were determined to have substantial overlap with other clusters, are shown in Supplementary Fig. S10 and S11.

To determine whether FRETfluors can perform consistently under a range of environmental conditions, we tested the sensitivity of FRETfluor spectroscopic signatures to changes in the local chemical environment by varying salt and pH across a physiological range (0 to 150 mM NaCl, pH 6.5-8.5). We found that the FRETfluor signals were consistent across all pHs tested, and exhibited small (~10%) salt-induced reductions in the brightness of $AB_{in}N$-type, AB*N*-type, and $A_cBN$-type constructs and in the donor lifetime of $AB_{in}N$-type constructs (Supplementary Fig. S12 and Supplementary Note S5). The shifts within each construct type might be amenable to calibration of a simple transformation so that FRETfluors can be accurately identified across variable environments. Moreover, the different responses across construct types indicate that the photophysical tuning of FRETfluors may be achieved by different combinations of local chemical and structural parameters, so that different constructs will exhibit greater or lesser sensitivity to the solution environment.

## Robust classification for mixtures of FRETfluors

We next sought to predict the maximum number of FRETfluors which could be reliably identified in a mixture, which depends on many factors including measurement duration and precision, as well as the set of molecules used for multiplexing and their potential for mis-classification within that set. To this end, we determined the largest set of FRETfluors for which the likelihood of misclassification between any individual pair was below 2.5%. From the pool of 41 FRETfluor constructs tested, we analyzed all possible pairwise combinations to determine which pairs presented higher chances of mutual misclassification. We fitted each cluster of levels from ABEL data as a three-dimensional Gaussian distribution in red and green brightness and green lifetime and performed a one-tail integration over the data parameter space to determine the likelihood of each FRETfluor being mis-identified as any other specific FRETfluor.

The matrix in Fig. 5a shows true cluster identity (left) and incorrect cluster identity (bottom), where each square is colored according to the probability of misidentification for that label combination. Any set of FRETfluors can only be used for unambiguous identification in the same sample if all its subset pairs have sufficiently low rates of misidentification. Here, the colorbar is capped at 2.5%, which we chose as a minimum criterion for this work to eliminate unfavorable label combinations. Supplementary Fig. S13 shows the same matrix on a full-scale colorbar (maximum misidentification probability: ~30% for $AB_{sk}10$ and AB8). Supplementary Fig. S14 shows the ranked order of pairwise misidentification scores, showing that few (33 of 1640 directional pairs) are above the 2.5% threshold and the likelihood of misclassification for most pairs is vanishingly small. From this analysis we identified a subset of 27 FRETfluors suitable for use in a single mixture (Fig. 5, arrows and Supplementary Table S3, bold) with the misclassification matrix shown in Supplementary Fig. S15.

Here, we allowed FRETfluors to exit the trap naturally, which typically occurred after a few seconds of trapping. Typical photon arrival rates are > 25 kHz summed across all channels, and level assignments are typically based on several thousand photons, sometimes with multiple levels per event (e.g., due to donor or acceptor blinking while trapped). To determine whether similar levels of discriminative power might be achieved with fewer photons, we examined the effect of the number of photons per point on the spread of FRETfluor clusters (Fig. 5b, Supplementary Note S6). For ~100 photons per point, clusters are broad and overlap, so that neighboring one- or two-tailed discrimination of a typical FRETfluor, AB11, from its closest neighboring populations only achieves 45% and 75% identification accuracy, respectively. However, with just 100 photons, AB11 can still be readily discriminated from nearby but non-neighboring clusters (AB8, 90% accuracy) or more distant clusters (AB16, 97.5% accuracy). With 3000 photons per point, identification is at worst 92% for two-tailed nearest neighbors and surpasses 97.5% at around 6000 photons per point for all cases tested. Thus, throughput of FRETfluor identification could be optimized by sampling each construct for only long enough to gather ~$10^4$ photons above background, which would take only 400 ms per construct (neglecting blinking effects).

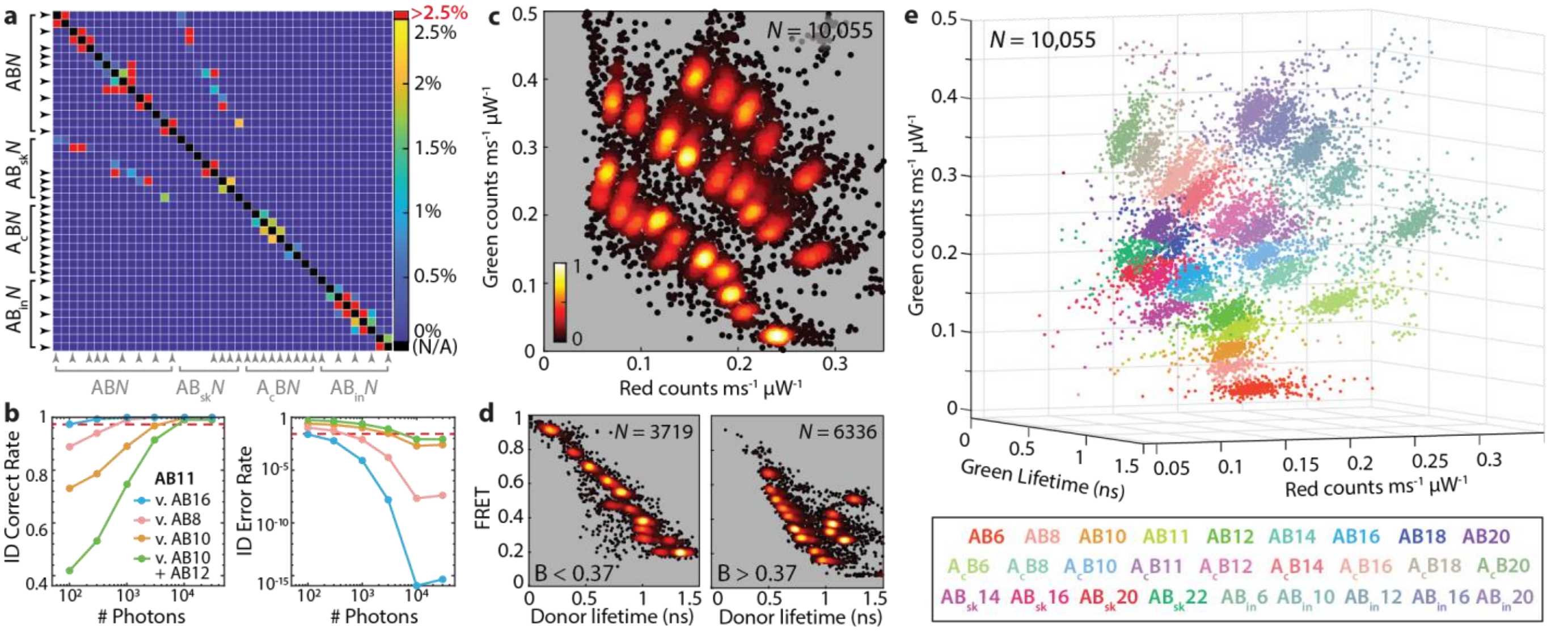

**Figure 5.** Selection and multiplexed detection of a near-orthogonal FRETfluor set. a) One-tailed Gaussian overlap between each pair of clusters was calculated for all 41 constructs. Colorbar represents the probability of misclassification for each pair, and is capped here at 2.5% (red). Self (correct) identification was not considered (black). b) Correct (left) and erroneous (right) identification rate of FRETfluor AB11 as a function of the number of photons considered, when considering confusion with either AB16 (blue), AB8 (pink), AB10 (orange), or both AB10 and AB12 (green). The cutoff of 97.5% is shown as a red dotted line. c) Red-green projection of data from multiplexed detection of 27 FRETfluors in a single sample. d) Two brightness slices of a lifetime-FRET projection, separated at B = 0.37 counts ms$^{-1}$ μW$^{-1}$. e) 3-D projection of spectroscopic data; each of 27 FRETfluor labels produces a cluster. Points are colored by cluster membership per the key below.

We next tested our ability to distinguish this optimized FRETfluor set experimentally. We combined the complete set of 27 FRETfluors in a dilute sample mixture (~2 pM total; ~75 fM of each FRETfluor). Fig. 5c shows a red-green projection illustrating clear separation of clusters; these clusters can be further differentiated in lifetime-FRET projections (Fig. 5d), here showing two brightness cuts above and below a total brightness of 0.37 counts ms$^{-1}$ μW$^{-1}$. A 3-D view captures these differentiated cluster positions within the detection space (Fig. 5e). Each cluster is produced by one FRETfluor label and is colored according to its most likely identity. Within this three-dimensional space, each of the 27 FRETfluor populations is easily distinguishable as predicted by Fig. 5a (rotating view of Fig. 5e is provided as Supplementary Video S1). The raw data shown in Supplementary Fig. S2 uses this combination of FRETfluors and is annotated with tag identities. Supplementary Fig. S16 shows the location (oval: 95% confidence interval) for each tag in this mixture, colored according to the construct type.

## Applications to detecting low-abundance targets in biomolecular mixtures

To explore the suitability of FRETfluors for applications requiring multiplexed detection of low-abundance targets, we tested both simple and complex mixtures of mRNA, dsDNA, and proteins (target biomolecules are depicted in Fig. 6a). We first tested wash-free labeling and readout of a 1:1 mixture of two mRNAs, FLuc and EGFP. FRETfluor AB6 was targeted to luciferase mRNA and AB12 was targeted to EGFP mRNA using the bridge strands $BR_{FLuc}$ and $BR_{EGFP}$, respectively. FRETfluor AB10 with an off-target bridge was added to control for nonspecific labeling. All three labels were mixed with a 1:1 ratio of FLuc:EGFP mRNA. Analysis of the resulting ABEL trap data clearly shows three spectroscopic populations in the expected locations corresponding to AB6, AB10, and AB12 (Fig. 6b scatter heatmap). Separation of bound and unbound populations in each cluster revealed that both AB6 and AB12 bound their target mRNA with 71% and 73% binding efficiency, respectively, while AB10 had only an unbound population and therefore no cross-reactivity with AB6 and AB12 (Supplementary Fig. S17). Superimposing the bound populations for AB6 and AB12 on a scatter plot of all unbound populations (Fig. 6b inset) confirms that the spectroscopic signatures of the FRETfluors are unchanged by target binding.

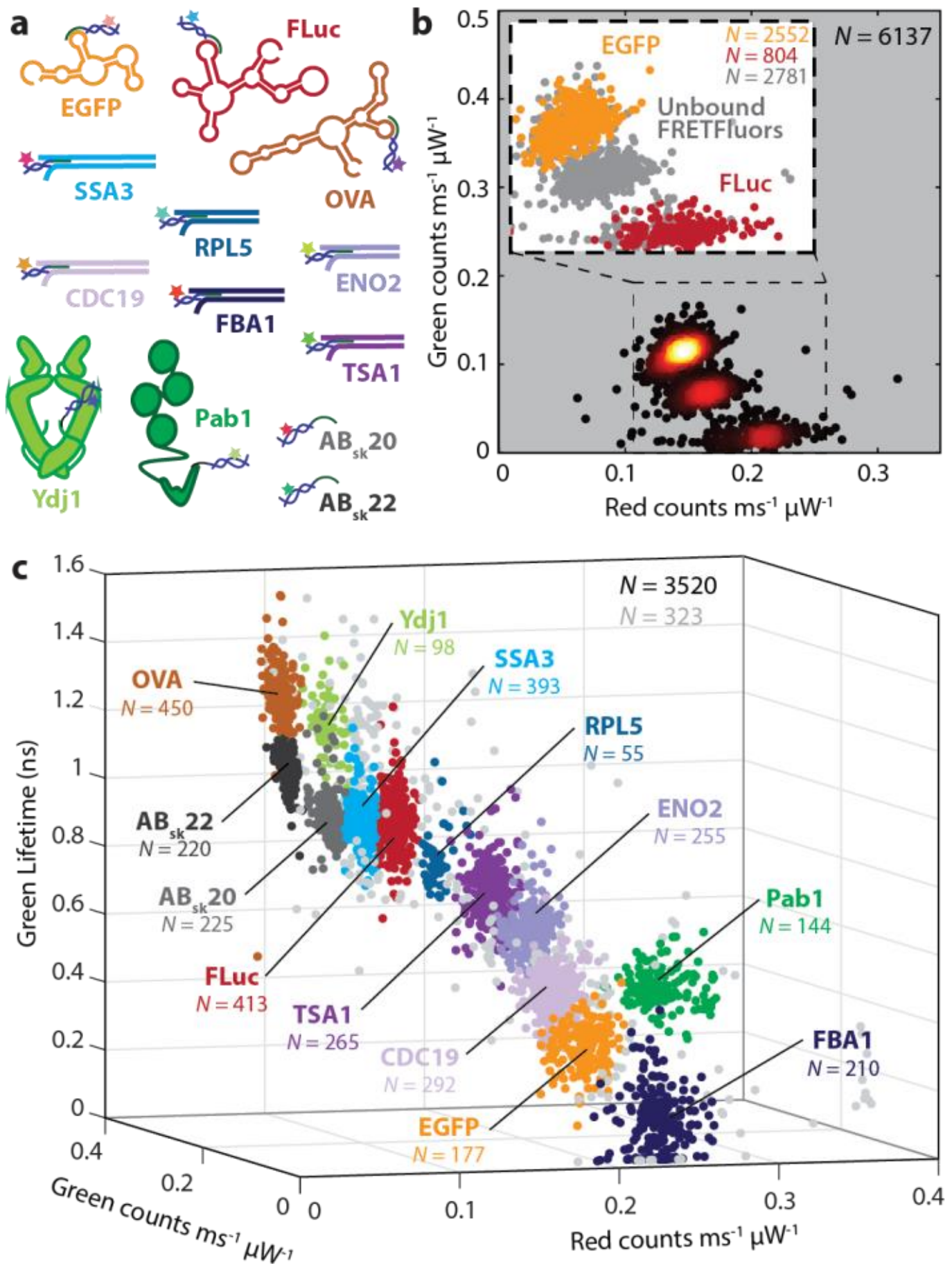

**Figure 6.** FRETfluor application to detection of complex mixtures of biomolecules at low concentration. a) Identity of all mRNA (3), dsDNA (6), proteins (2), and target-less FRETfluor (2) samples. b) Scatter heatmap of red-green projection of data for a simple mixture of EGFP+AB12, FLuc+AB6, and AB10 (no target). *Inset:* Scatter plot of signals from unbound FRETfluors of all types (gray) and bound EGFP+AB12 (orange) and FLuc+AB6 (red). c) 3-D projection of spectroscopic data for a complex mixture of mRNA, dsDNA, and proteins (cluster colors per Panel 6a), with unassigned levels shown as light gray points.

Finally, we tested FRETfluor multiplexed detection of many different species and types of molecules present in low concentrations in a complex sample, as might be found in biomedical or environmental samples. We separately labeled mRNAs (EGFP, FLuc, and OVA), proteins (Pab1 and Ydj1), and dsDNA fragments that were RT-PCRed from abundantly expressed and stress response-related transcripts in *S. cerevisiae*[63] (FBA1, CDC19, ENO2, TSA1, RPL5, and SSA3), as detailed in Supplementary Tables S4 and S5. A subset of the 27-FRETfluor combination tested above, selected to generate closely neighboring clusters (AB6, 8, 10, 11, 12, 14, 16, 18, and 20, along with $AB_{sk}16$ and $A_cB6$) were used for labeling with appropriate targeting bridge strands. Together with two FRETfluors with off-target bridges as controls ($AB_{sk}20$ and $AB_{sk}22$), labeled targets were mixed in approximately equal proportions and diluted to a total working concentration of 5 pM (~350 fM each). Analysis of the resulting ABEL trap data shows all FRETfluor labels present and with their expected spectroscopic signatures (Fig. 6c); clusters do not shift upon binding a target. Closer examination of each cluster (Supplementary Fig. S18) shows that bound and unbound FRETfluors can be differentiated as expected, and that the two off-target FRETfluors do not cross-react with other elements of the mixture. These results provide a promising proof-of-concept for use of FRETfluors in applications that require characterization of dilute, highly heterogeneous mixtures of many different types of biomolecules.

## Discussion

Our approach to designing FRETfluor labels goes against most conventional wisdom for creating FRET constructs: here we *utilize* the sensitivity of fluorophores to their local environments and attachment chemistries, including local sequence, altering photophysical properties to enhance multiplexing without requiring additional chemical structures. We observed that tuning dye properties allows the characteristic emission properties of entire sets of FRET constructs to be shifted to previously unused sectors of the multi-parameter readout space. Tuning dye properties therefore acts as an additional dimension that expands the design space for engineering protein fluorophore-sized constructs with distinctive spectroscopic emission.

We also showed that relatively small changes in dye photophysics are sufficient to produce uniquely identifiable sets of FRETfluors. The $AB_{sk}N$ constructs are only 15-20% different from the AB*N* constructs in donor brightness and lifetime, yet generate well-separated states. Critically, for different donor attachment chemistries and sequence contexts, we observed that Cy3-only brightness and lifetime did not change in a perfectly correlated way, indicating that the radiative lifetime must be changing alongside the non-radiative lifetime.[49] This is particularly useful for FRETfluor design: Supplementary Notes S7 and S8, along with Supplementary Fig. S19, detail additional simulations of FRET on DNA showing that decoupled donor lifetime and brightness lead to nearly orthogonal directional shifts of a FRET curve within the multi-parameter space used in this work. Although not addressed in this work, shifts in the acceptor properties should have similar potential to expand the number of identifiable constructs. We expect that development of more precise control over individual fluorophore properties will only further expand multiplexing possibilities with the approach described here.

Additional unique advantages of this approach are conferred by using the ABEL trap for sample readout, in particular its capabilities for extended duration multi-parameter measurements and for estimation of sample transport properties. Here, we were able to reliably multiplex classification of single molecules at ultra-low concentration (tens of femtomolar per species). We did not attempt to optimize throughput in this work; the ABEL trap measures only one molecule at a time, and we did not limit the time spent on each molecule. For higher throughput, a statistically optimal number of photons necessary to correctly identify each species could be determined, and feedback could be turned off after that predetermined time for each detected event, with a small bias voltage to bring new sample into the sensing region. Parallelized microfluidic readout channels and spectroscopic detection could permit additional throughput gains. A related challenge will be quantitatively connecting levels and events measured to the stoichiometry of a mixture. In the present work, we found that cluster populations were approximately, but not quantitatively, reflective of the presumed mixture stoichiometry. This discrepancy may arise from complicating factors such as trapping bias or dye photophysics, which might influence both the number and proportion of observed levels meeting the analysis filter criteria. We expect that this effect could in principle be calibrated out for each sample and label combination.

We show here that FRETfluors are particularly useful for wash-free sensing at ultra-low concentrations, but such conditions also necessarily present labeling efficiency challenges. When labeling at ultra-low concentrations, a high proportion of detected FRETfluors will be free rather than bound to a target molecule, lowering the rate of useful detected signals. It will be necessary to consider tradeoffs between labeling efficiency and throughput on an application-by-application basis. We used the exquisite sequence specificity of nucleic acid hybridization and the site-specificity of cysteine labeling on proteins to prevent cross-reactivity in the applications shown here, but we expect that other common targeting methods such as antibody recognition will bring their own optimization challenges in terms of efficiency, potential cross-reactivity, and label size.

Cyanine dyes are well-known to be sensitive to their environment in contexts beyond DNA attachment, for example via protein induced fluorescence enhancement (PIFE)[64,65] or solution and local environment composition.[36,66] Here, PIFE was not observed upon FRETfluor attachment to proteins, nor were the FRETfluor signals altered by binding to mRNA or DNA. We also observed that FRETfluor signals were robust to a limited range of physiological pH, and that certain constructs exhibited slight shifts with changing salt. Together, these results suggest that FRETfluor structure and attachment chemistry are the dominant environmental influences on cyanine dye photophysics in this study, and that the FRETfluor structure may partially protect the cyanine dyes against interactions with target molecules or solvent. Use of non-isomerizing[67] or otherwise photostabilized[68] cyanine dyes could further protect FRETfluors from environmental effects. Regardless, FRETfluor signals will need to be carefully calibrated for each application's potential target and solution environment. We envision selecting FRETfluor sets with clusters spaced far enough apart to allow for small shifts due to potential environment or target effects. If, as our current data suggests, construct families exhibit similar environmentally-induced shifts, these patterns could prove useful to maintain cluster separability even under changing conditions. Moreover, it may be the case that for certain applications, sensitivity of a FRETfluor signal to the target or environment will confer sensing functionality; for example, detecting a change in the target's composition, confirming attachment to a low molecular-weight target that does not change trapping characteristics, or providing additional information about the local environment composition. Ultimately, the range of FRETfluor tuning that is possible with any particular dye chemistry will depend upon the mechanisms by which its photophysics can be tuned by the local construct structure and chemistry, balanced with the effects of the solution environment.

The approach demonstrated here is in principle cross-compatible with many of the elegant strategies for single-molecule spectroscopic multiplexing that have been proposed by others, including additional excitation lasers, dye ratios, or colors,[4,3,34] orientational control of dyes to influence polarization,[27,44] making use of the full distributions available across the multiple detected parameters[23] rather than simplified or averaged values as we show here, as well as with barcoding-types multiplexing strategies such as DNA-PAINT.[7,12,14] Other molecular scaffolds or fluorophore types could be used to create FRETfluors, and scaffold geometries and sizes could be altered and optimized. Other readout parameters, including dye orientation, might provide additional tuning dimensions. We expect that FRETfluors could also be used in a wide-field imaging format, given sufficiently sensitive wide-field FLIM capabilities, although a wide-field approach would not capture the size information necessary for wash-free detection. Finally, here our analysis and classification are based on average values of levels in the data, but we expect that the rich data collected for FRETfluors would lend itself to more sophisticated analyses involving machine learning, which could take full advantage of the information present in the data to extract higher-fidelity classification of tag identity and target binding status.

The primary technological advance of FRETfluors is the use of tunable dye photophysics as an additional multiplexing dimension complementary to, and therefore combinatorially expanding, the multiplexing power of FRET. The advantages of multiplexing with a minimal set of chemical components are both practical and technical: Fewer components improve simplicity and could reduce costs, while also providing chemical consistency across constructs to achieve robust performance over diverse environment, and allowing for additional modifications to further enhance multiplexing. By combinatorially expanding the degree of multiplexing that is possible with a limited set of building blocks, FRETfluors open up previously inaccessible design space for multiplexing applications at the single-molecule level.

# Methods

**DNA oligo samples and FRETfluor preparation**

All oligos were purchased with fluorophores from IDT and purified by high-performance liquid chromatography. Full sequences for all oligos are given in Supplementary Table S1. Most oligos include either /iCy3/ or /iCy5/ internal modifications (Cy3 and Cy5 are non-sulfonated). The labeling efficiencies of each strand were 70%-90% by absorption measurements. Double-stranded DNA constructs were annealed by mixing the complementary strands at 5µM concentration in TE buffer (pH 8.0), heating to 90 °C for 2 minutes, and slowly cooled down to 25°C with steps of 0.5°C per 20 seconds. DNA samples were stored at 4°C prior to use. Bulk fluorescence emission was characterized with Fluorolog®-3 with FluorEssence™ in shared facilities at the University of Chicago. Bulk fluorescence lifetime measurements for Cy3 were taken with a ChronosBH with magic angle detection and confirmed via bulk measurements on the ABEL setup. For bulk characterization, samples were excited with a 520 nm laser to minimize direct excitation of the acceptor.

**Single-molecule characterization in the ABEL trap**

A custom Anti-Brownian ELectrokinetic (ABEL) trap was constructed after Ref. 52, incorporating excitation with a pulsed (60MHz) supercontinuum laser (Leukos ROCK 400-4) paired with an AOTF (Leukos TANGO VIS) to output 532 nm excitation light. The excitation spot was steered in the sample plane using two AODs (MT110-B50A1.5-VIS) arranged orthogonally and driven by a direct digital synthesizer (DDSPA2X-D8b15b-34). Fluorescent photons are split into red and green channels using a dichroic filter (Chroma T610lpxr). Each channel is then split into s- and p-polarized light using a polarizing beam splitter (Thorlabs) and focused onto separate APDs (Excelitas SPCM-AQRH-14-ND, four total). For each detected photon, a TCSPC (Picoquant Multiharp 150) records the time and the color and polarization channel in which it arrived.

APD signals are also sent to an FPGA (NI PCIe-78656) which controls the ABEL trap. Based on the position of the laser upon the arrival of each fluorescent photon, taking into account a pre-calibrated lag, the position of the fluorescent molecule is estimated via a noise-rejecting Kalman filter.[53] XY voltages proportional to the estimated displacement of the trapped particle relative to the center of the trap are passed to a 10x voltage amplifier (Pendulum F10AD). The amplified voltages are applied to a quartz microfluidic sample cell via platinum electrodes which sit in four reservoirs at the cardinal points of the microfluidic cell.

The microfluidic cells are cleaned and passivated prior to each trapping experiment. Microfluidic cells were cleaned in piranha solution (3:1 mixture of sulfuric acid and hydrogen peroxide) overnight and then rinsed extensively with ultra-pure DI water. The microfluidic cells were incubated in 1 M KOH for 10 minutes and then passivated using mPEG-silane (Laysan Bio MPEG-SIL-5000, 50 mg $ml^{-1}$) in 95% ethanol, 5% water with 10 mM acetic acid for 48 hours.[34] The cells were rinsed with ultra-pure DI water and incubated with 1% Tween[69] for 10 minutes and rinsed thoroughly before adding the sample for measurements.

Immediately prior to measurement, FRETfluors were diluted to a total concentration of between 1-5pM in buffer containing 20 mM HEPES (pH 7.4), 3 mM Trolox, and an oxygen scavenging system (~60 nM protocatechuate-3,4-dioxygenase and 2.6 mM protocatechuic acid).

**Analysis of ABEL trap data**

All data analysis was performed using customized software written in Matlab, after Ref. 52. Briefly, photon arrival times recorded by the Multiharp were used to construct a 10 ms-binned time trace, for which the background levels were identified for each channel using an information-criteria-optimized (AIC) K-means clustering algorithm. Raw photon arrival timestamps were used to identify brightness change points in each channel using the algorithm of Watkins and Yang,[69] which were merged into a single list of change points. Data between each pair of change points was assigned to one "level".

For each level, background-subtracted brightness in all four detection channels was determined and used to calculate a FRET value. Note that here we use both FRET and donor lifetime parameters only for the purpose of separating distinct spectroscopic signals, so that further corrections are not necessary for the purpose of this analysis. To facilitate comparison of our data with data generated by other labs, we include the FRET correction parameters for our setup in Supplementary Note S9. For donor lifetimes, we used the green parallel channel only, fitted by maximum likelihood on iterative convolution with the IRF to a single exponential decay for each level, after Zander et al[70] and Brand et al[71] as previously described.[52,22] IRFs were collected using a short lifetime fluorescent dye (malachite green). Although 2- and 3-exponential decays provide better fits to the data, for the purpose of this work and given the low photon count in many levels, we observed more robust label classification with 1-exponential fits for lifetime data. Data from levels of duration greater than 150 ms were used for subsequent analysis and can be viewed as individual points in figure scatter plots.

For certain analyses, levels were further binned into groups of $M$ photons (statistical separation of clusters based on $M$ photons, calculation of $x$-$y$ position fluctuations within trap); remainder photons at the end of each level were unused.

**Determination of FRETfluor classification clusters and statistical overlap**

K-means clustering in three dimensions (red and green brightness, donor lifetime) was used for initial classification of events from each data set involving more than one FRETfluor label. The mean and standard deviation in each cluster was determined using a 3-D Gaussian fit without covariance to the cluster after rejecting all outliers (>3σ from initial cluster mean). Means and standard deviations from the same FRETfluors across different data sets collected on different days were compared to verify cluster consistency.

Statistical pairwise overlap of these normalized 3-D Gaussians (green brightness, red brightness, green lifetime) was used to calculate the probability of pairwise misclassification, defined as the summed probability in the overlapping tails of each pair of distributions.

**Nucleic acid labeling**

Bridge strand sequences (BR) for targeting all ssDNA, mRNA, and dsDNA, including null (non-targeting) sequences, as well as the amine-modified bridge sequence for site-specific protein labeling, are given in Supplementary Table S1. FRETfluor constructs were incubated with the ssDNA target during the annealing process, with binding confirmed by EMSA. Polyacrylamide gels (12%) for ssDNA binding were run at 200 V in TBE.

We used mRNA for enhanced GFP (EFGP), ovalbumin (OVA) and firefly luciferase (FLuc), all commercially available (CleanCap mRNA, Trilink BioTechnologies). Incubation with mRNA samples was carried out at 37 °C for 18 hours in labeling buffer (TE pH 8.0) containing RNase inhibitor (10 U/µL, SUPERase•In™ RNase Inhibitor). After annealing, both electrophoretic mobility shift assay (EMSA) and ABEL trap data confirmed mRNA binding. EMSA for mRNA was done following a standard protocol at room temperature,[72] on agarose gels (3%) run at 110 V in 0.5x TAE. Max current was set to 50 mA.

To generate dsDNA targets for FRETfluor binding, total RNA was extracted from wild-type BY4742 yeast (*S. cerevisiae*) grown in YPD medium to OD 0.3-0.4, using a Direct-zol RNA purification kit (Zymo Research) with on-column Dnase I treatment for at least 15 minutes. The purified RNA was reverse transcribed (iScript Select, Bio-Rad) with custom gene-specific primers (IDT; sequences are given in Supplementary Table S5). The combined RNA and cDNA was further amplified by conventional PCR (Q5 Hot Start Master Mix, New England Biolabs) using asymmetric priming to favor production of the coding DNA strand (1 uM forward primer, 25 nM reverse primer). Final target DNA amplicons ranged from 516 to 1262 bases in length. The RT-PCR products were then hybridized with the designed bridge by heating to 90 °C for 2 minutes and slowly cooled down to 25°C with steps of 0.5°C per 20 seconds. Targeting bridges were designed to attach to the 25 bases next to the forward primer region of the target sequence to avoid competing against free primer. For target labeling, FRETfluor tags hybridized to the appropriate bridge strand were introduced and the mixture was incubated at room temperature for 3 hours. 3% agarose gels were run at 110V in 1x TAE, and target bands were cut and soaked in TE buffer (pH 8.0) overnight for extraction. After a clarifying spin (10,000 g, 2 minutes), 150 µL of the supernatant was removed as the final dsDNA-FRETfluor stock sample.

**Protein labeling**

FRETfluors were conjugated to proteins via a bifunctional small molecule linker (Pierce™ SMCC, No-Weigh™ Format, Thermo Fisher) containing both an NHS ester group and a maleimide group. FRETfluors were pre-hybridized with a bridge DNA strand modified at the 5' end with a primary amine ($BR_{amine}$). The assembled FRETfluor was reacted with linker at an approximately 500:1 molar ratio (linker to FRETfluor) in phosphate buffered saline pH 7.4 (Gibco) for five hours at room temperature, covered, with shaking at 300 rpm. The FRETfluor-linker mixture was then buffer exchanged into fresh buffer to separate unreacted linker from FRETfluor (Zeba 7K MWCO, Thermo Fisher).

In parallel, target proteins mutated to have a single accessible cysteine (*S. cerevisiae* Pab1 C70A/C119A/C368A/A577C and Ydj1 C29A/C370F) were reduced in 5 mM TCEP for 45 minutes at room temperature, then buffer exchanged into reducing agent-free phosphate buffered saline pH 7.4 (Gibco) by spin column (Zeba 7K MWCO, Thermo Fisher). Protein concentration after buffer exchange was measured by absorbance at 280 nm (NanoDrop One).

The reduced protein and FRETfluor-linker samples were then combined in an approximately 100:1 molar ratio (protein monomer to FRETfluor) and incubated for two hours at room temperature, covered, with shaking at 300 rpm.

The protein-linker-FRETfluor mixture was mixed with 50% v/v glycerol to a final concentration of 10% glycerol and loaded onto a polyacrylamide gel (Mini-PROTEAN® TGX™ 4–20%, Bio-Rad). Samples were electrophoresed at 200 V for 20 minutes in Tris/glycine/SDS running buffer. Two reference lanes (Pab1, Ydj1) were excised and stained with Coomassie (Gel Code Blue, Thermo Fisher) to visualize protein. The gel was pieced back together and imaged (ChemiDoc, Bio-Rad) in Cy5 and Cy3 fluorescence channels to determine the location of FRETfluors on the gel. Gel fragments (unstained) containing protein and bound FRETfluor but not free FRETfluor were excised and incubated in the dark at 30°C with shaking for 10 hours with 500 µL gel extraction buffer (50 mM Tris-Cl, 150 mM NaCl, 0.1 mM EDTA, pH 7.5). After 10 hours, the buffer was separated from the gel pieces and subjected to a clarifying spin (10,000 g, 10 minutes). 300 µLof the supernatant was removed as the final protein-FRETfluor stock sample and stored at 4°C.

**Sample preparation for mixture applications:**

After preparing all FRETfluor-labeled mRNA, dsDNA, and protein samples individually, fluorescence correlation spectroscopy was used to estimate each sample concentration. The ABEL trap setup (beam not scanned, feedback off) was used to collect this data. Samples were diluted to 100-500 fM each to generate a concentration of ~5pM total sample in buffer containing 20 mM HEPES (pH 7.4), 3 mM Trolox, and an oxygen scavenging system (~60 nM protocatechuate-3,4-dioxygenase and 2.6 mM protocatechuic acid).

## Acknowledgment

The authors acknowledge valuable discussions and feedback from: Natalie Tsang, Jared Bard, Samantha Keyport Kik, David Pincus, Justin Jureller, Juan de Pablo, and members of the Squires Lab. The authors thank the Bukau Lab for their generous donation of *H. sapiens* HSPA8 for use as a reference reagent. K.M.L. acknowledges support from NIH F30 1F30ES035279-01, T32GM007281-45, and the Grier Prize from the UChicago Institute for Biophysical Dynamics. D.A.D. acknowledges support from NIH award GM144278. A.H.S. acknowledges support from the Neubauer Family Foundation as well as NSF QLCI QuBBE grant OMA-2121044 and seed funding from NSF MRSEC grant DMR-2011854.

## Author contributions

J.C., A.E., K.M.L., and A.H.S conceived and designed constructs and experiments. J.C., A.E., K.M.L, and M.R.J. performed experiments. A.E. and A.H.S. constructed the ABEL trap, and J.C. fabricated microfluidic measurement cells. K.M.L. and D.A.D. provided additional reagents and samples. J.C., A.E., K.M.L., and A.H.S. analyzed data and created figures. A.E.C., J.C., A.E., and A.H.S. worked on simulations. J.C., A.E., and A.H.S. wrote the manuscript. All authors reviewed and edited the manuscript.

## Competing interests

None

## Materials and correspondence

Correspondence should be directed to asquires@uchicago.edu. The authors declare that the data supporting the findings of this study are available within the paper and its Supplementary Information files. Raw figure data files are available on Zenodo [XXX IN PREPARATION; LINK TO BE INSERTED HERE]. Oligo sequences for all unique materials created in this work are detailed in Supplementary Table S1 and Supplementary Table S5 and are commercially available.

# SUPPLEMENTARY INFORMATION:
# Single-molecule fluorescence multiplexing by multi-parameter spectroscopic detection of nanostructured FRET labels

**Jiachong Chu[1*], Ayesha Ejaz[2*], Kyle M. Lin[3,4], Madeline R. Joseph[1], Aria E. Coraor[1], D. Allan Drummond[5,6,7], and Allison H. Squires[1,7⌂]**

[1]Pritzker School of Molecular Engineering, University of Chicago, IL, USA
[2]Department of Chemistry, University of Chicago, IL, USA
[3]Graduate Program in Biophysical Sciences, University of Chicago, Chicago, IL, USA
[4]Interdisicplinary Scientist Training Program, Pritzker School of Medicine, University of Chicago, Chicago, IL, USA
[5]Department of Biochemistry and Molecular Biophysics, University of Chicago, Chicago, IL, USA
[6]Department of Medicine, Section of Genetic Medicine, University of Chicago, Chicago, IL, USA
[7]Institute for Biophysical Dynamics, University of Chicago, Chicago, IL, USA
*Authors contributed equally to this work

⌂e-mail correspondence: asquires@uchicago.edu

# Table of Contents

# Supplementary Figures

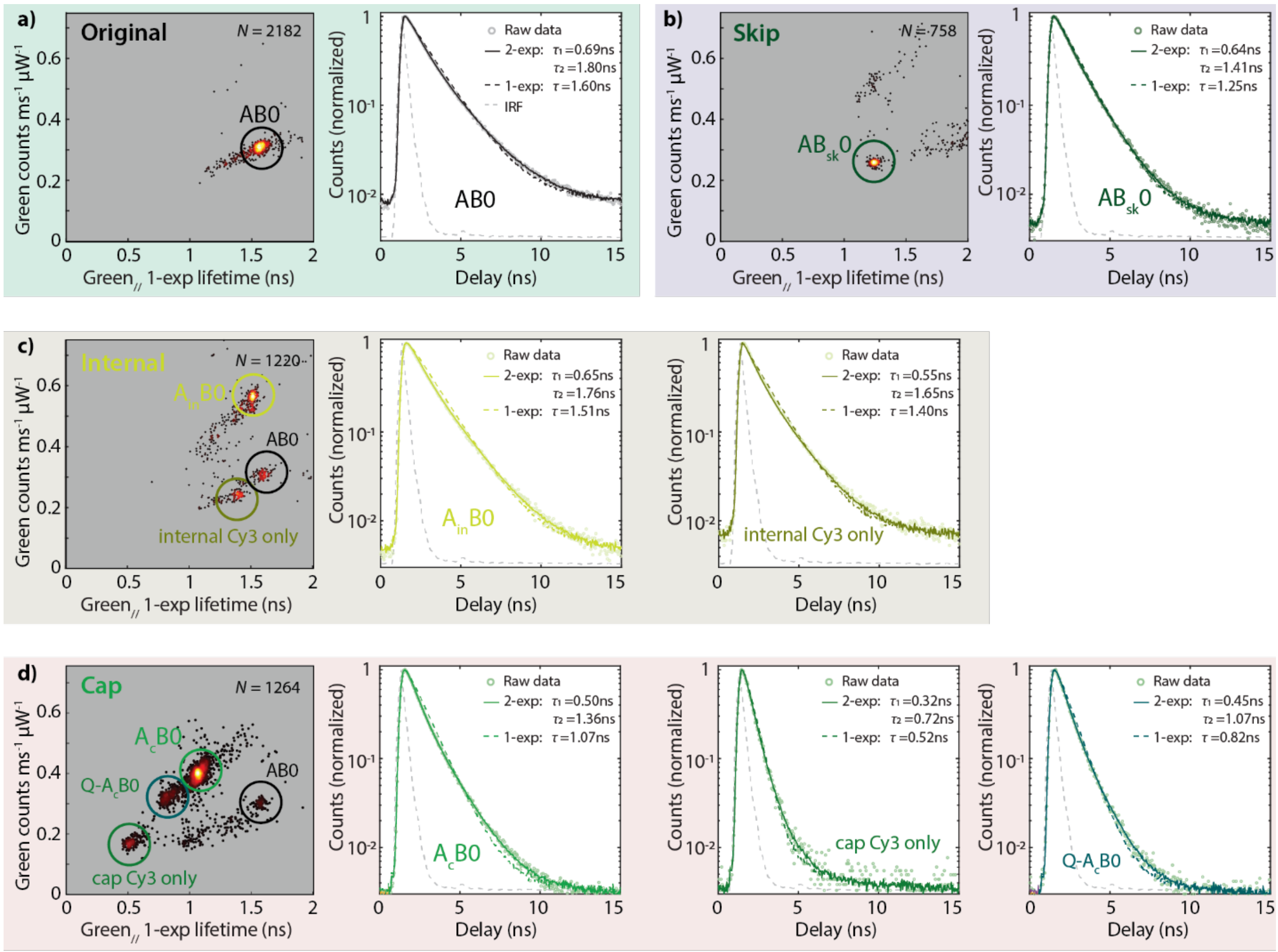

Figure S1. Single-molecule brightness and lifetime data for donor-only constructs. a) Levels for construct AB0 show one cluster. b) Levels for construct $AB_{sk}0$ also show just one population with a slightly shorter lifetime and dimmer brightness. Note that the changes in brightness and lifetime are not proportional to one another. c) Levels for construct $A_{in}B0$ show three populations: the original AB0 population, a population for the internal Cy3 label, and the majority population when both Cy3s are on. d) The levels for $A_cB0$ constructs show multiple states, of which the brightest population has both Cy3s on (as verified by allowed transitions into and out of this state). The AB0 population is unchanged. The cap-Cy3-only population can be deduced to be the dim, short-lifetime population. The fourth population is unknown. In all cases, the 2-component lifetime fits are superior, but the 1-component lifetime fits adequately represent the weighted average of the better fits. IRFs are shown as gray dashed curves.

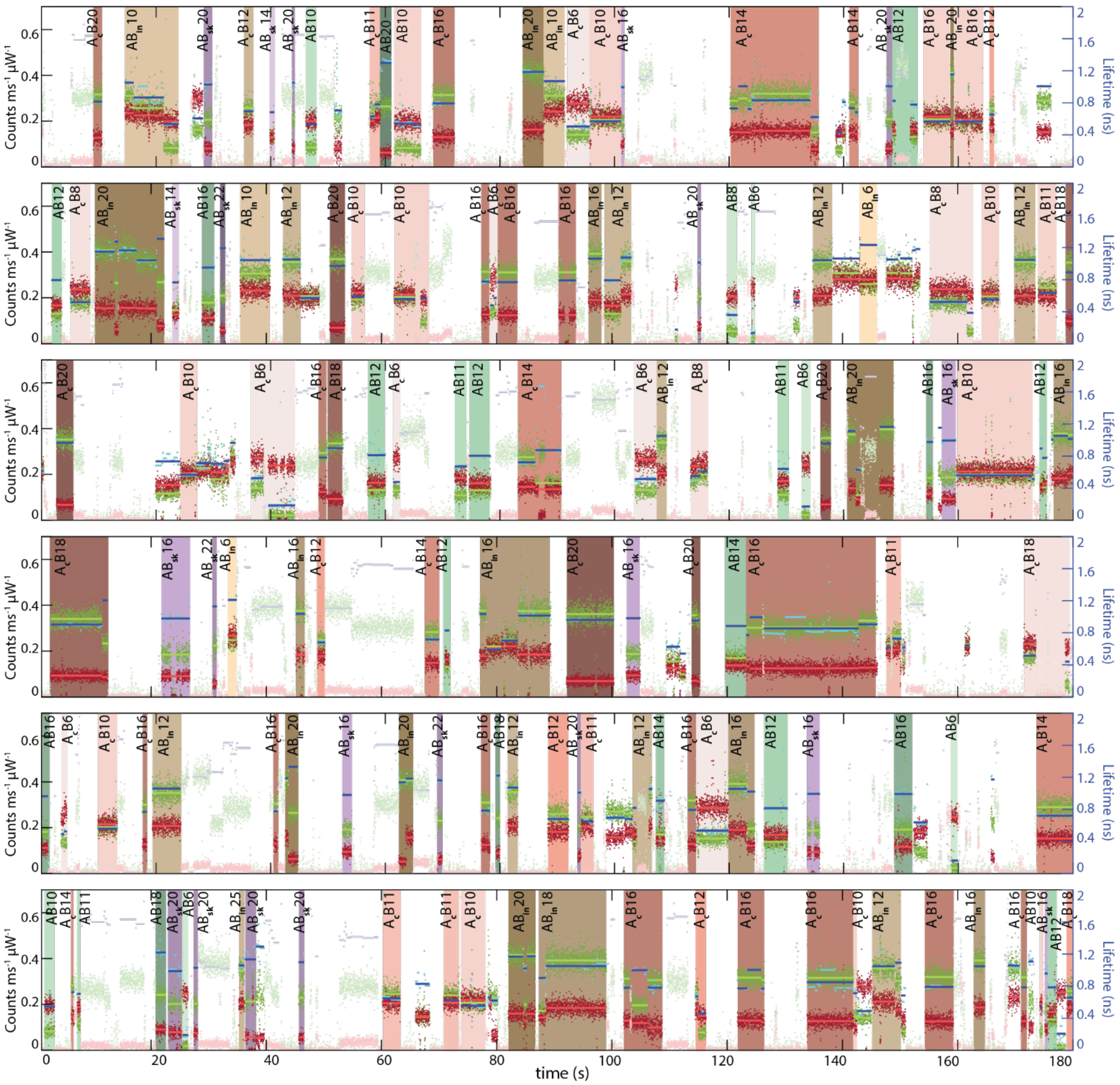

Figure S2. Extended raw data trace for ABEL trapping of single tags. Raw trapping data is shown for green brightness, red brightness, and green channel lifetime for 27 unique FRETfluor tags over a representative 18-minute data set. After events were filtered out when the red channel brightness is lower than 0.5 counts $ms^{-1}$ $\mu W^{-1}$, the rest of the events were identified according to the most likely cluster identity from Supplementary Table S3, with a limit of 1 standard deviation from the mean. Most unclassified (dim) events are excluded because they are donor-only. Highlight colors and FRETfluor labels correspond to the populations illustrated in Supplementary Fig. S16 and detailed in Supplementary Table S3. Desaturated data indicates a donor-only event; saturated levels without colorblocks indicate an event that could not be classified.

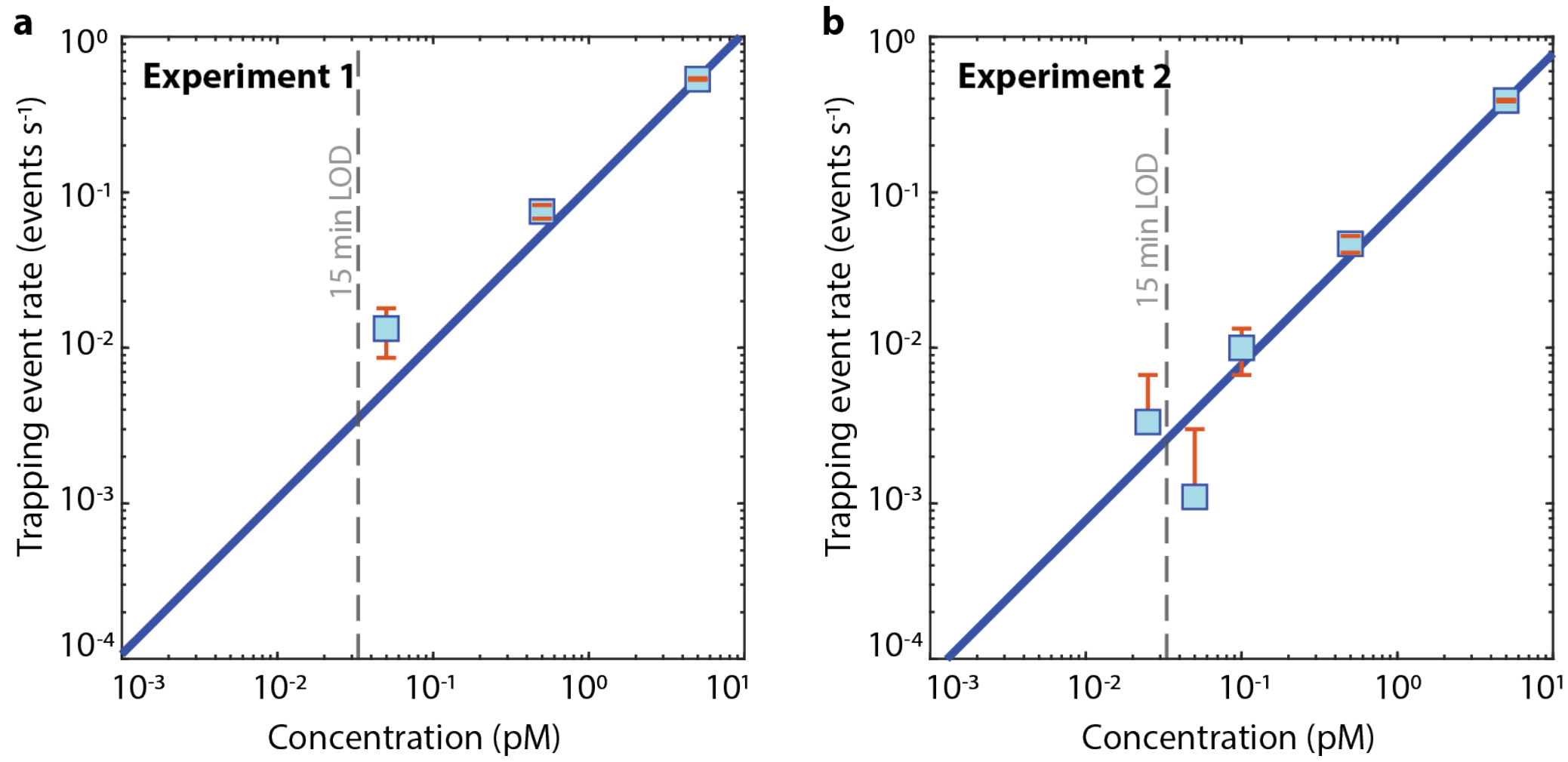

Figure S3. Limit of detection for FRETfluors in the ABEL trap. a) Observed trapping event rate for AB12 at increasing concentration (50 fM, 1 pM, 5 pM; 3 data sets of 5 min each; fit slope 0.08 events $s^{-1}$ $pM^{-1}$). b) Observed trapping event rate for AB12 at decreasing concentrations (5 pM, 500 fM, 100 fM, 50 fM, 25 fM; 3 data sets of 5 min each; fit slope 0.1 events $s^{-1}$ $pM^{-1}$). Gray dotted lines indicate the calculated limit of detection (LOD) for a 15-minute measurement, ~33 fM (see Supplementary Note S3 for details).

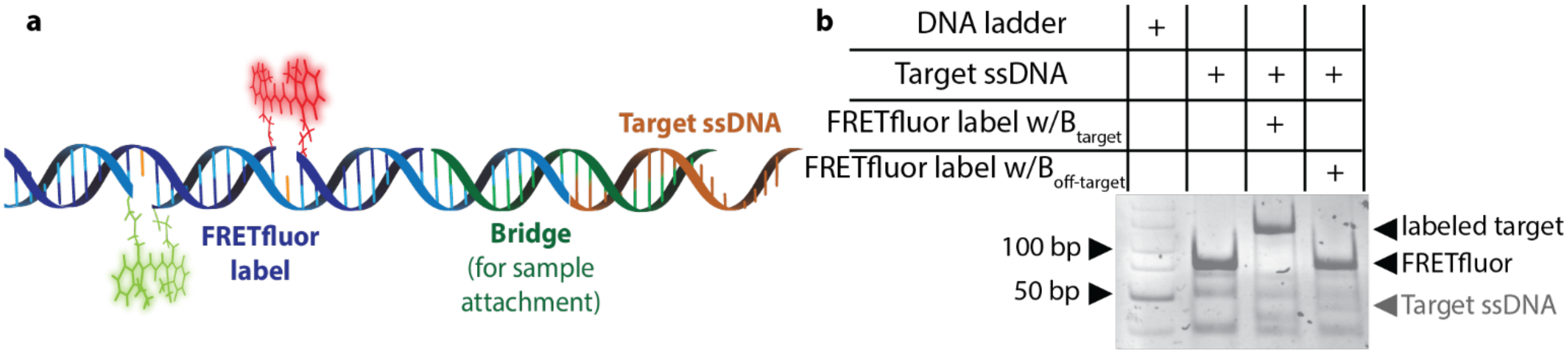

Figure S4. Binding to ssDNA. a) Illustration of a FRETfluor construct hybridized to a ssDNA target sequence (blue: FRETfluor tag, green: labeling bridge, orange: target DNA). b) EMSA showing a mobility shift for on-target binding to a 55bp ssDNA (lane 3) as compared to free label (lane 2), but no shift for an off-target bridge sequence (lane 4).

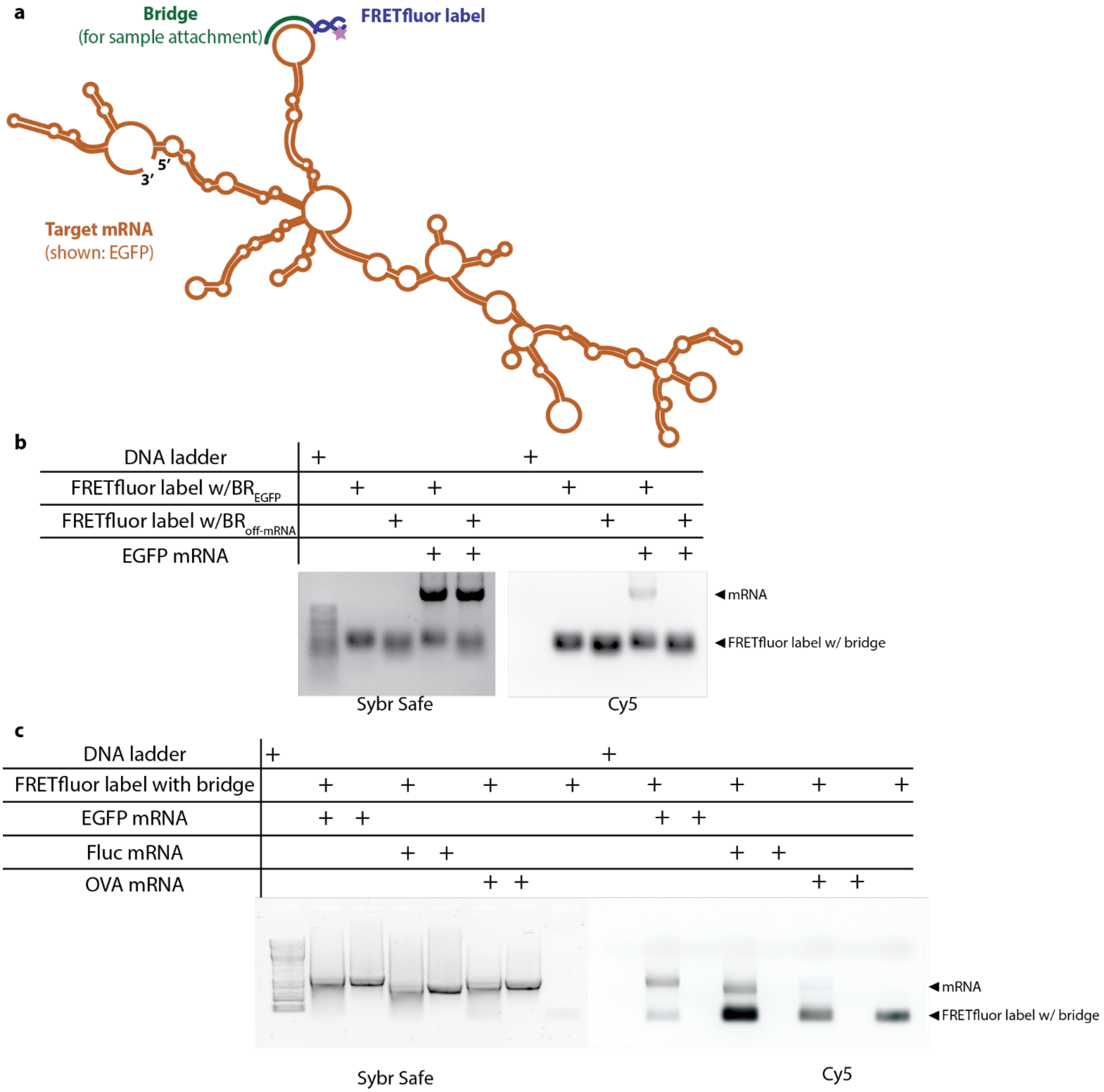

Figure S5. FRETfluor binding to mRNA. a) Illustration of a FRETfluor label (blue) attached to a predicted hairpin loop of the EGFP mRNA through a labeling bridge (green). b) Electrophoretic mobility shift assay (EMSA) on an 1% agarose gel showing a mobility shift for on-target binding to an EGFP mRNA as compared to free label (lanes 2 and 3), but no shift for a mismatched labeling bridge. The gel was scanned in two different excitation channels: Trans UV for SyberSafe (left) and Cy5 excitation channel (right). c) Electrophoretic mobility shift assay (EMSA) on an 1% agarose gel showing a mobility shift for on target binding to EGFP mRNA, FLuc mRNA and OVA mRNA as compared to the free tag itself (lanes 2, 4, 6 and lane 8) but no shift for the target itself (lanes 3, 5, 7). The gel was scanned in two different excitation channels: Trans UV for SYBR Safe (left) and Cy5 excitation channel (right).

**a**

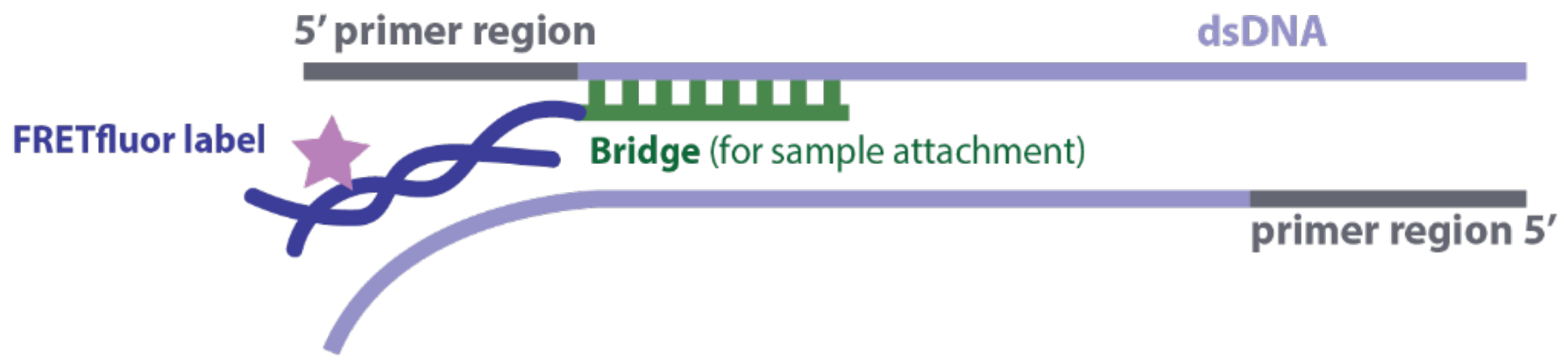

**b**

| | | | | | | | | |
|---|---|---|---|---|---|---|---|---|
| DNA ladder | + | | | | + | | | |
| FRETfluor label | | + | + | + | | + | + | + |
| PDC1 RT-PCR product | | + | | | | + | | |
| FBA1 RT-PCR product | | | + | | | | + | |

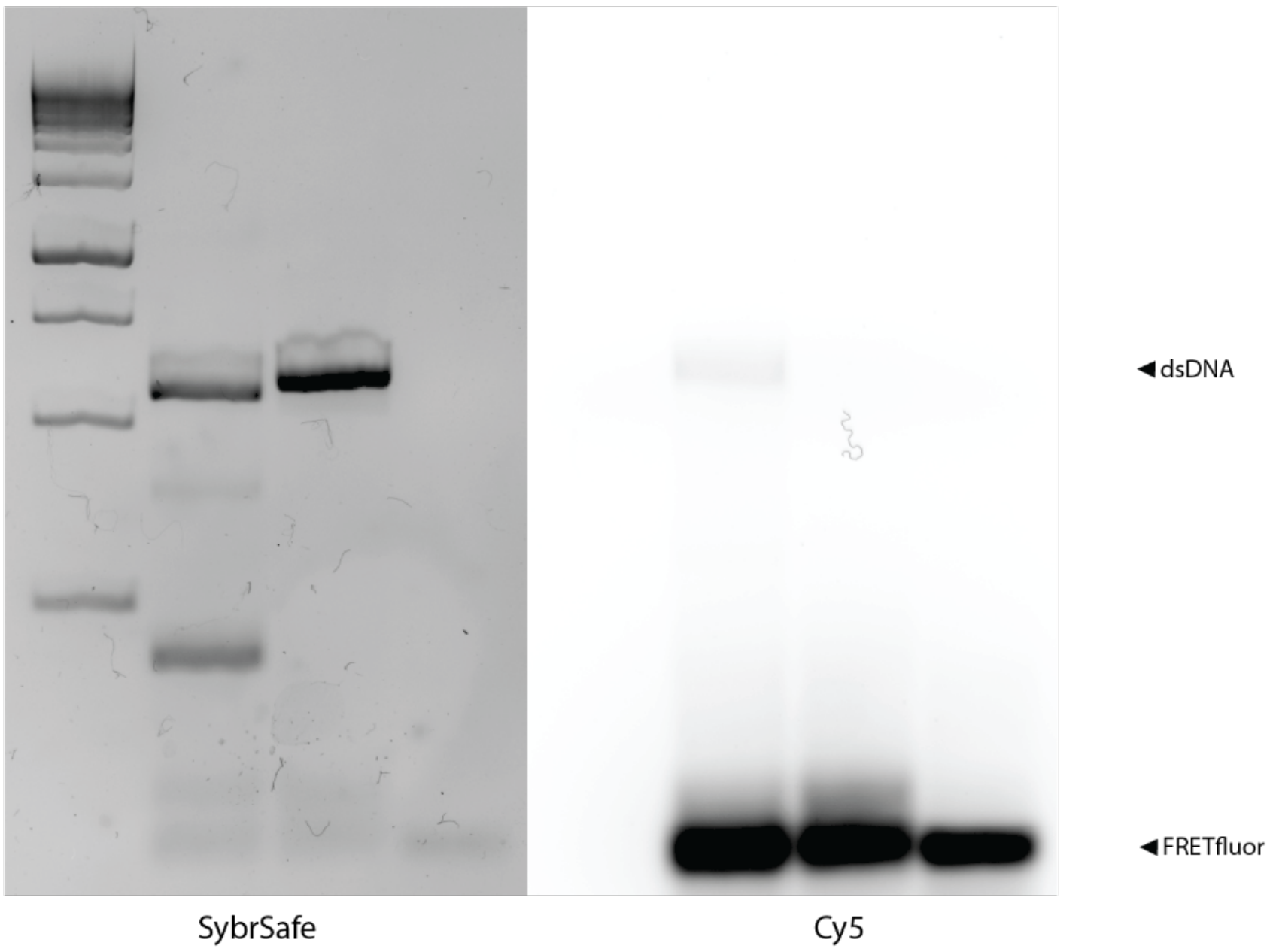

Figure S6. FRETfluor binding to dsDNA. a) Illustration of a FRETfluor label (blue) attached to a 25 base-long targeted region of dsDNA, with a 3-base gap from the primer region on the dsDNA product (purple) through a labeling bridge (green). b) In this assay, the bridge was designed to be complementary to the PDC1 RT-PCR dsDNA. Electrophoretic mobility shift assay (EMSA) on a 3% agarose gel shows a shift upon target binding to the correct dsDNA (lane 2). No binding or shift occurs for the wrong target (lane 3), confirming labeling specificity. The gel was scanned in two different excitation channels: Trans UV for SYBR Safe (left) and Cy5 excitation channel (right).

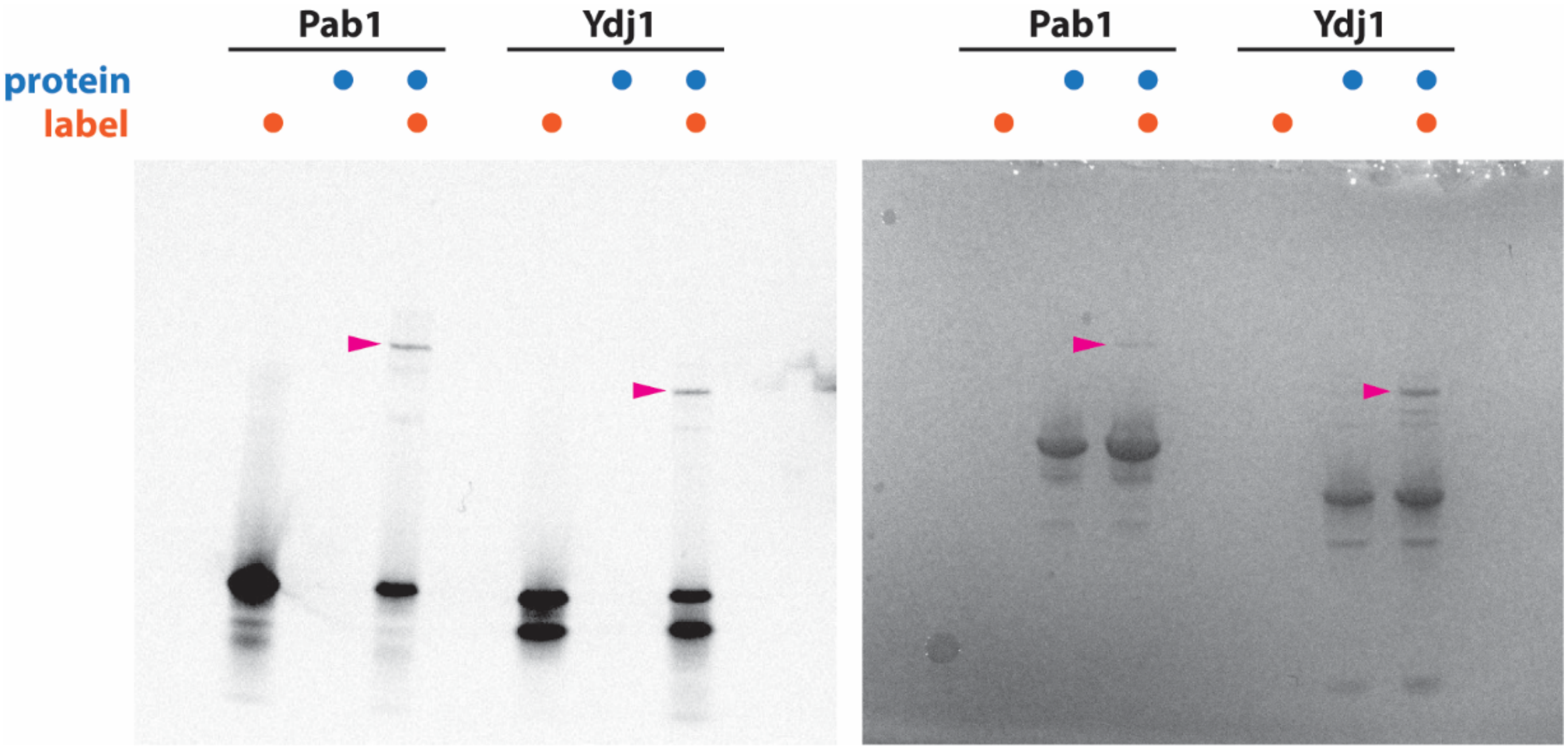

Figure S7: FRETfluor binding to proteins. Two different single-cysteine mutant proteins (Pab1 C70A/C119A/C368A/A577C and Ydj1 C29A/C370F) reacted with FRETfluors via a covalent NHS ester-maleimide linker, as well as protein and FRETfluor alone, were compared by polyacrylamide gel electrophoresis and imaged in Cy5 fluorescence channel (left) or with white light after protein staining (right). In each protein-FRETfluor reaction, an upshifted band appears in the Cy5 and protein images (magenta arrows), indicating formation of the correct protein-DNA complex.

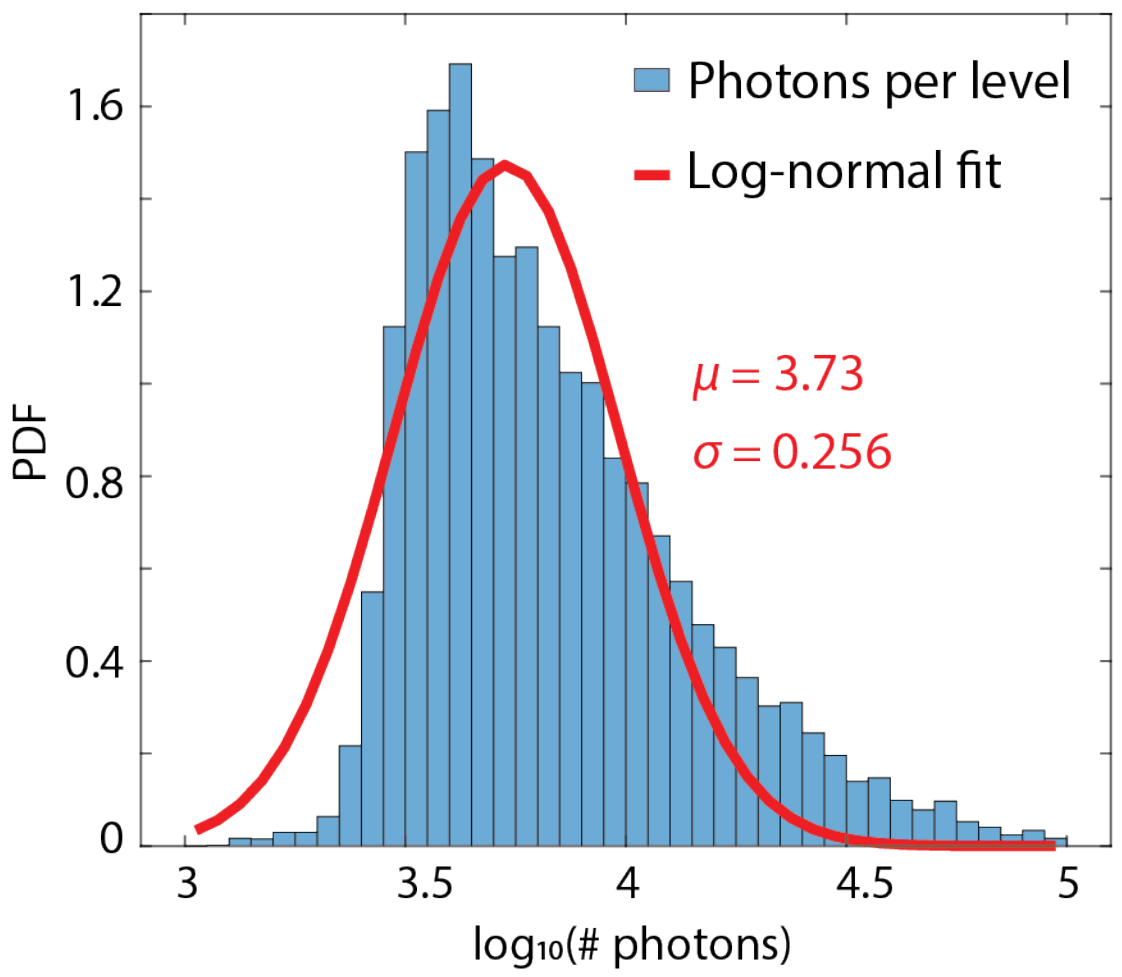

Figure S8: Photons per data level. Log-normal histogram and fit of the #photons per measured level for a mixture of FRETfluors (red and green channels combined, background subtracted). Mean of fit: $10^{3.73}$ = 5354 photons (-$\sigma$: 2971 phot, +$\sigma$: 9650 phot).

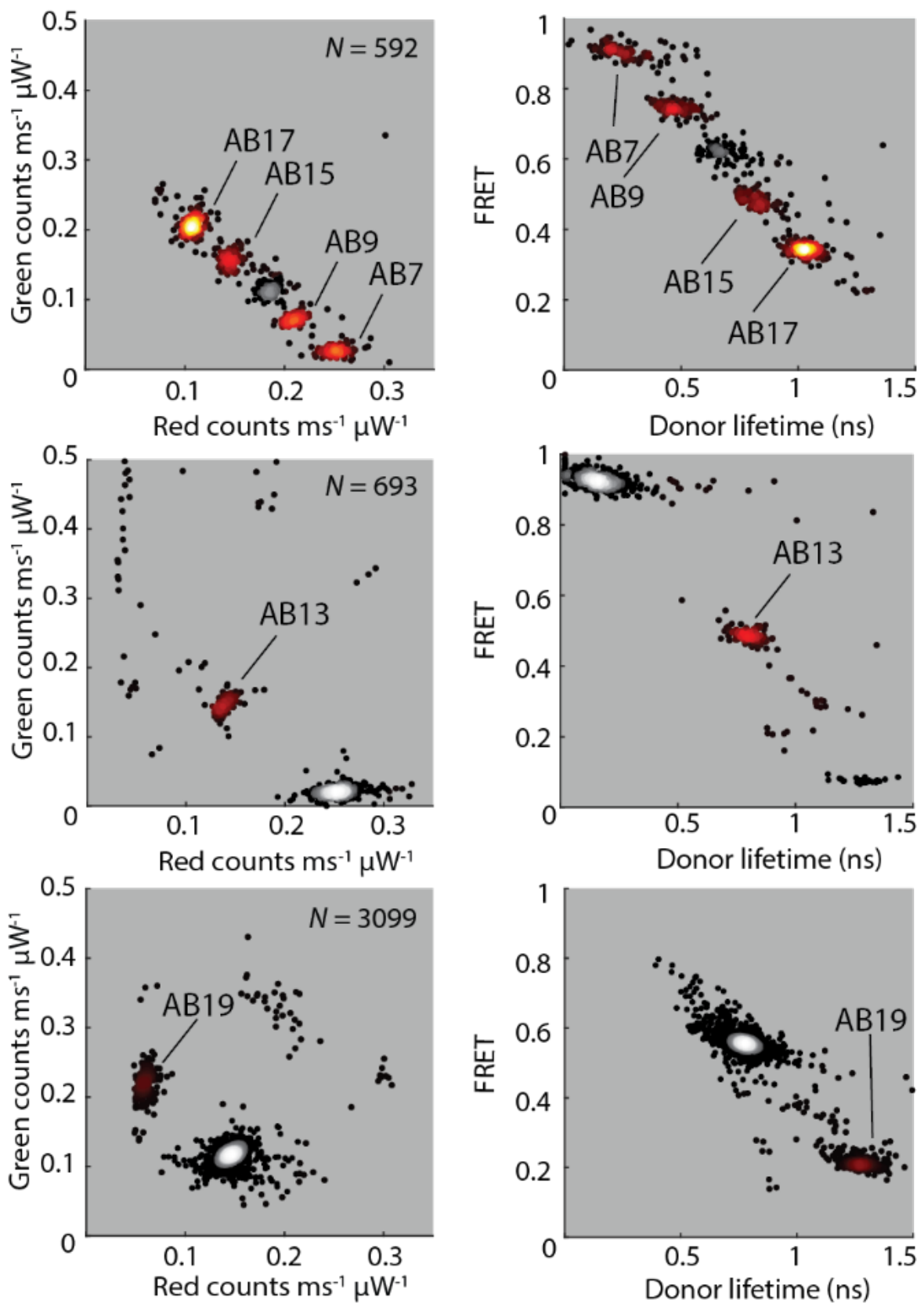

Figure S9. Additional constructs from the ABN series that are not shown in Fig. 4.: Red-green and lifetime-FRET projections of data from trapped FRETfluors with design variations show clusters in different regions of the parameter space.

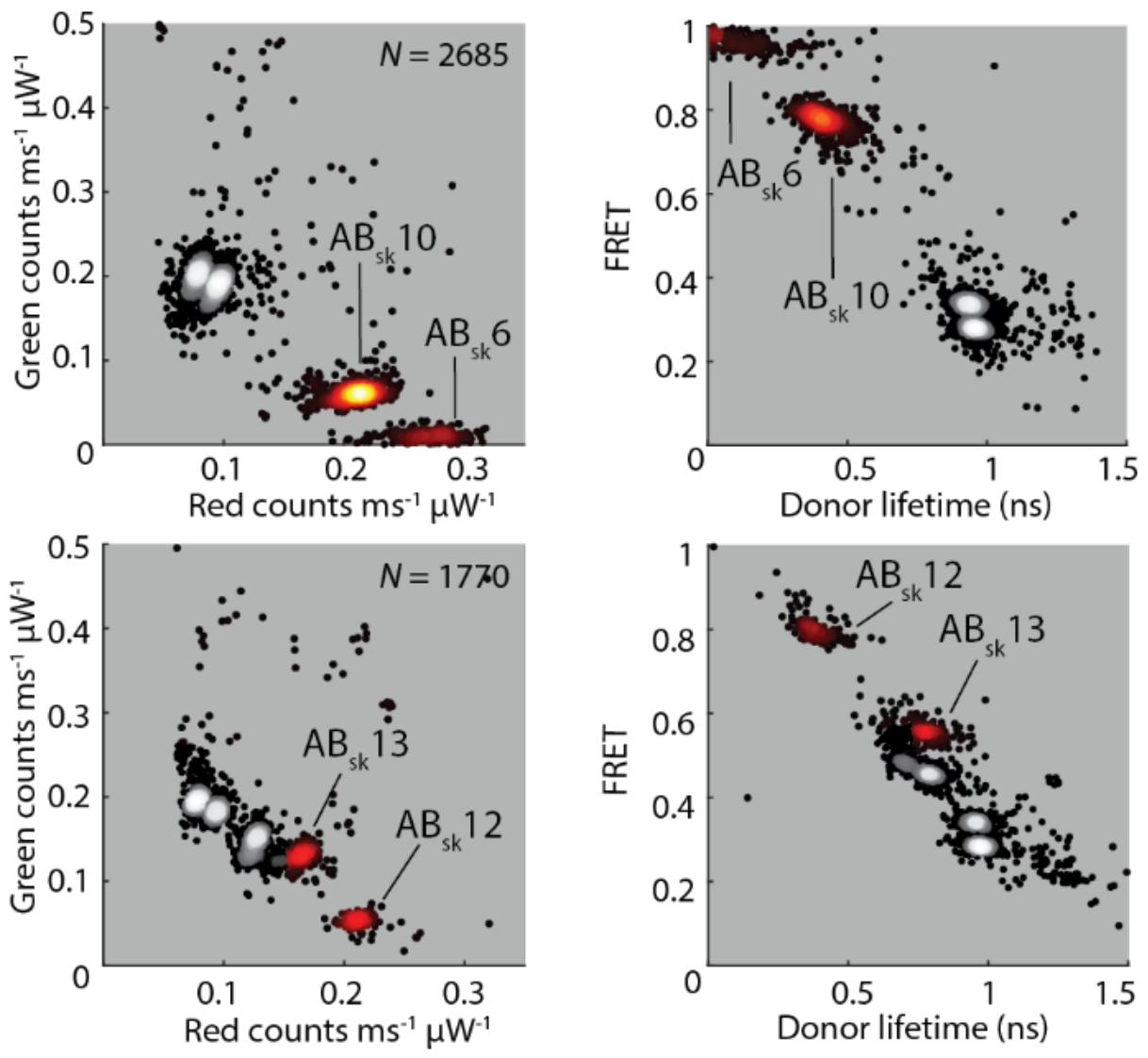

Figure S10. Additional constructs from the $AB_{sk}N$ series that are not shown in Fig. 4.: Red-green and lifetime-FRET projections of data from trapped FRETfluors with design variations show clusters in different regions of the parameter space.

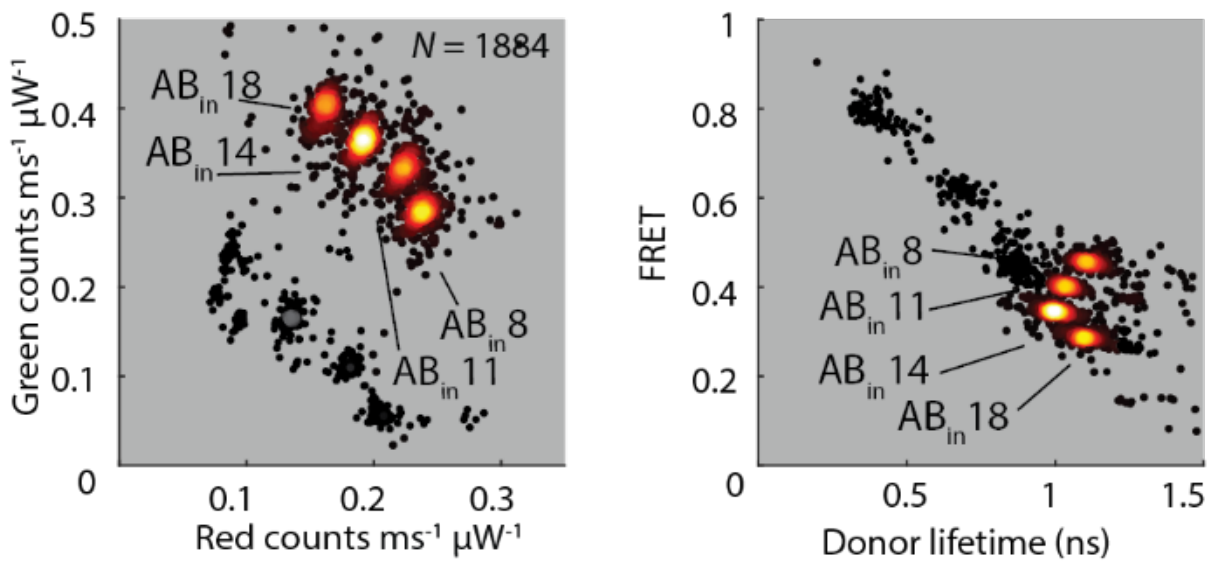

Figure S11. Additional constructs from the $AB_{in}N$ series that are not shown in Fig. 4.: Red-green and lifetime-FRET projections of data from trapped FRETfluors with design variations show clusters in different regions of the parameter space.

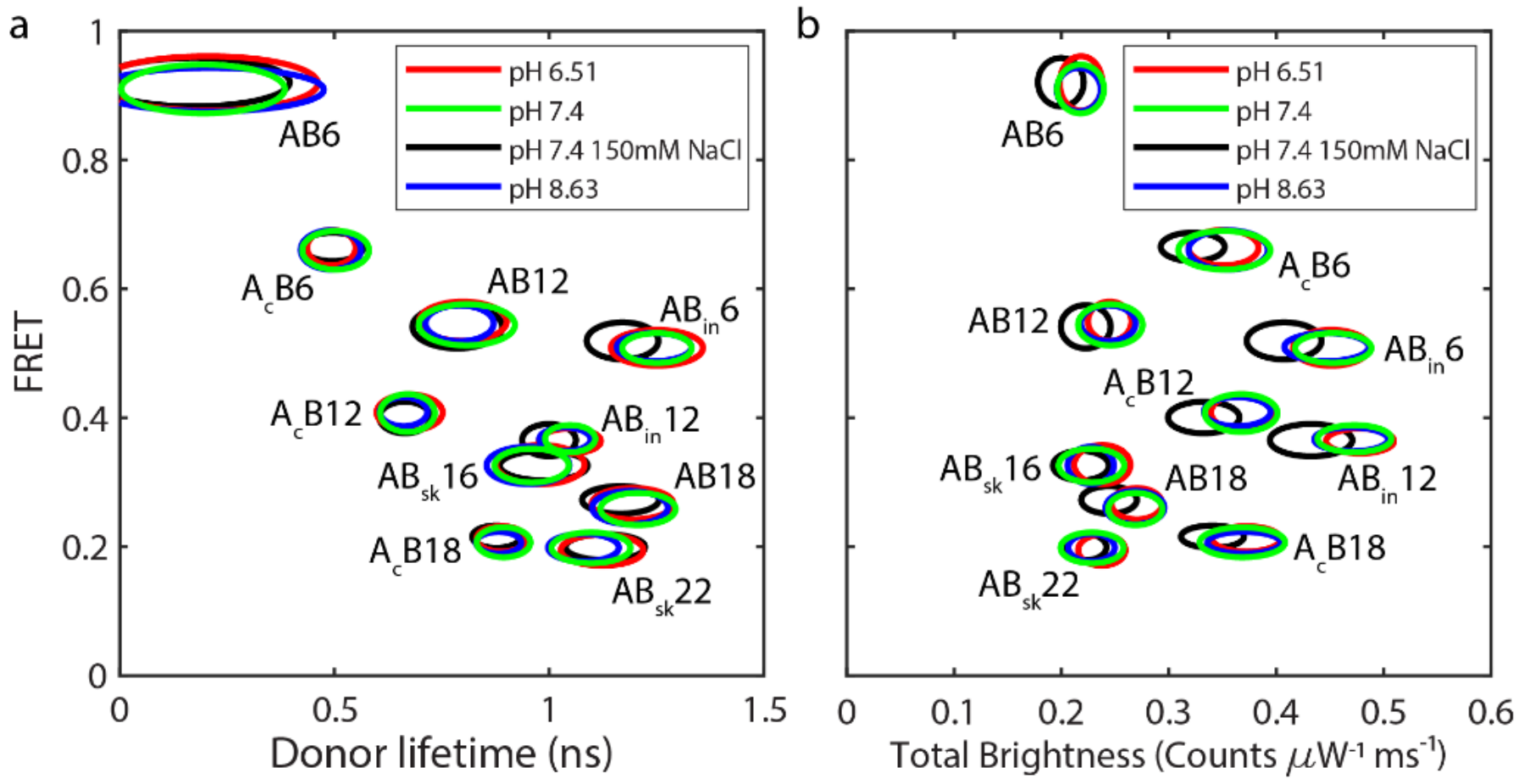

Figure S12. FRETfluors in different salt and pH conditions. a) FRET-donor lifetime projection and b) FRET-brightness projection of cluster locations for 10 representative FRETfluors of different construct types. Cluster properties are unchanged by pH across the range 6.5-8.6, but $AB_{in}N$-type clusters show slightly reduced donor lifetime in 150 mM salt, and $AB_{in}N$-type, $ABN$-type, and $A_cBN$-type clusters all show slightly reduced total brightness in increased salt. The $AB_{sk}N$-type constructs appear unchanged across all conditions.

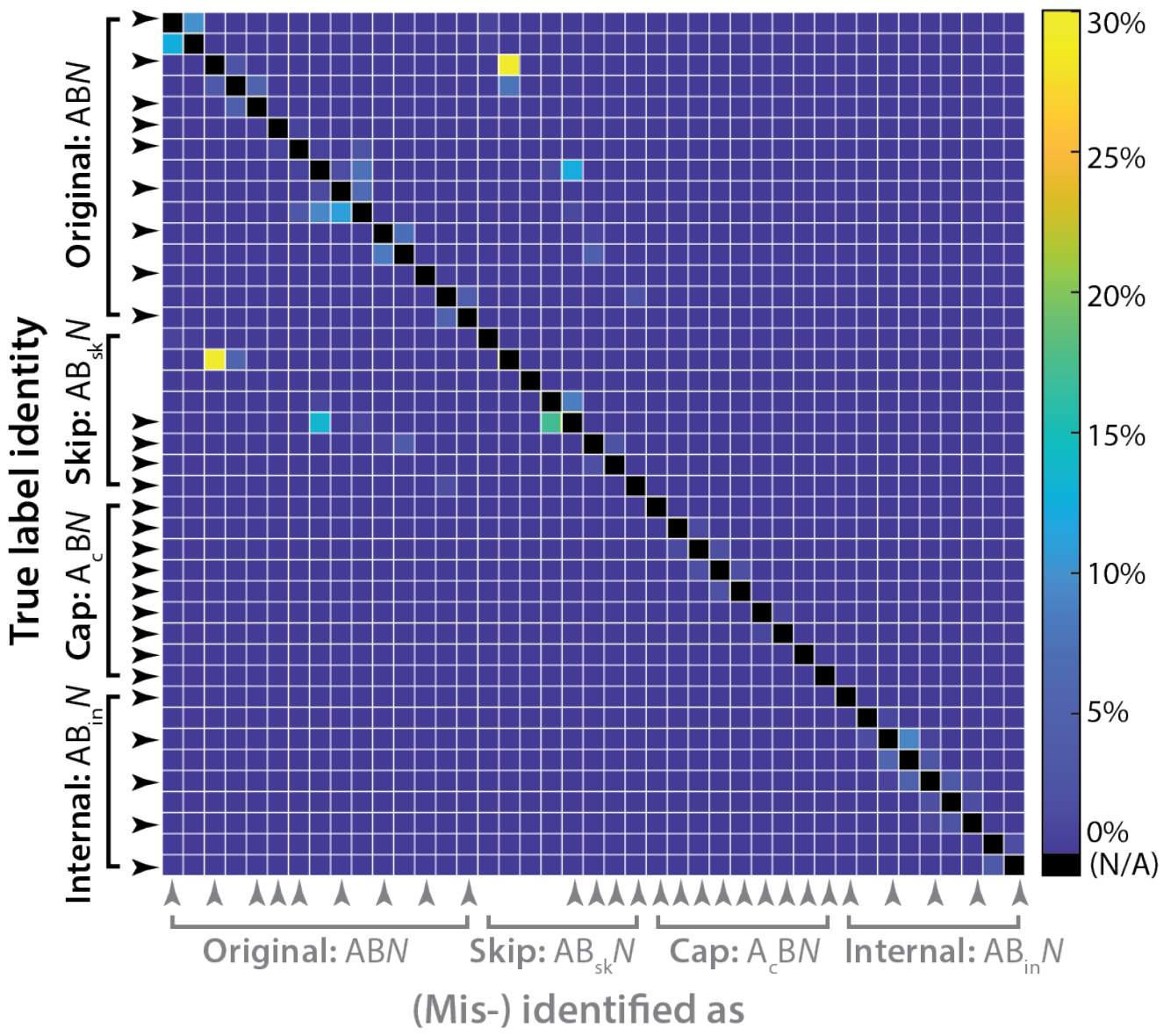

Figure S13. The detail of the pairwise distribution analysis. Statistical selection of a near-orthogonal FRETfluor set. One-tailed Gaussian overlap between each pair of clusters was calculated for all 41 constructs. The color bar cutoff is 30%.

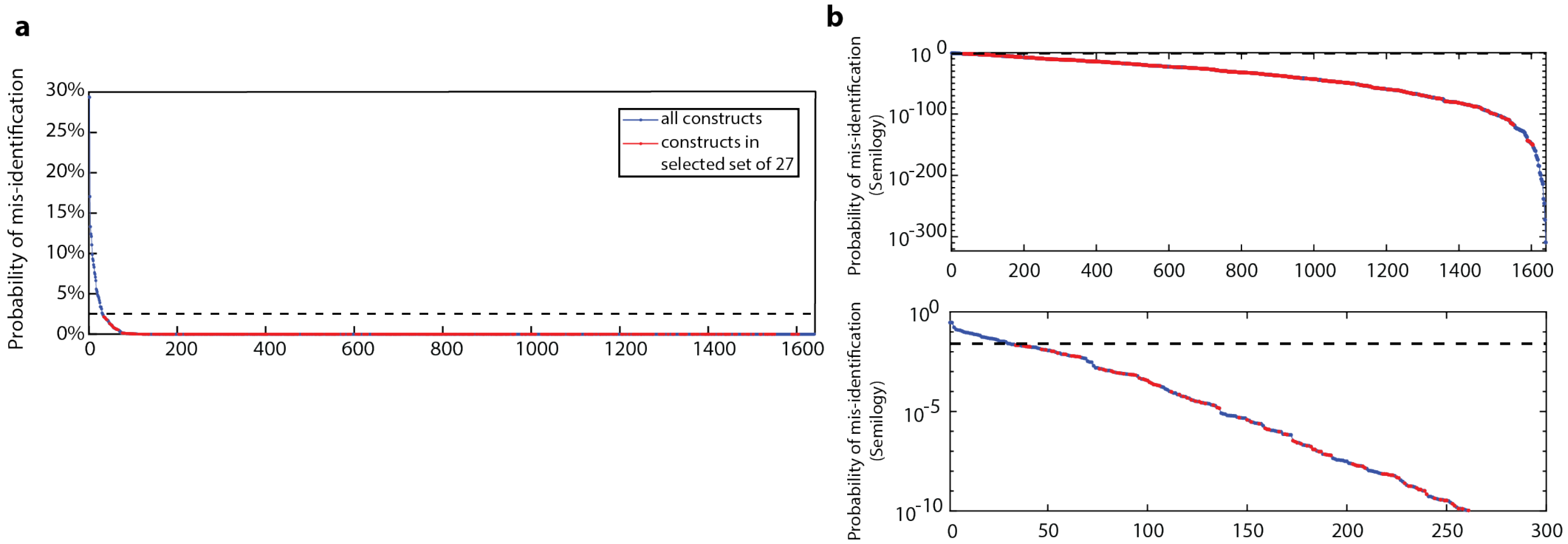

Figure S14. Pairwise misidentification distributions. a) The ranked probabilities of pairwise misidentification for all possible FRETfluor pairs. b) Same data, y-axis shown on a log scale. Top panel is the full plot, bottom is zoomed in to show detail at higher probabilities.

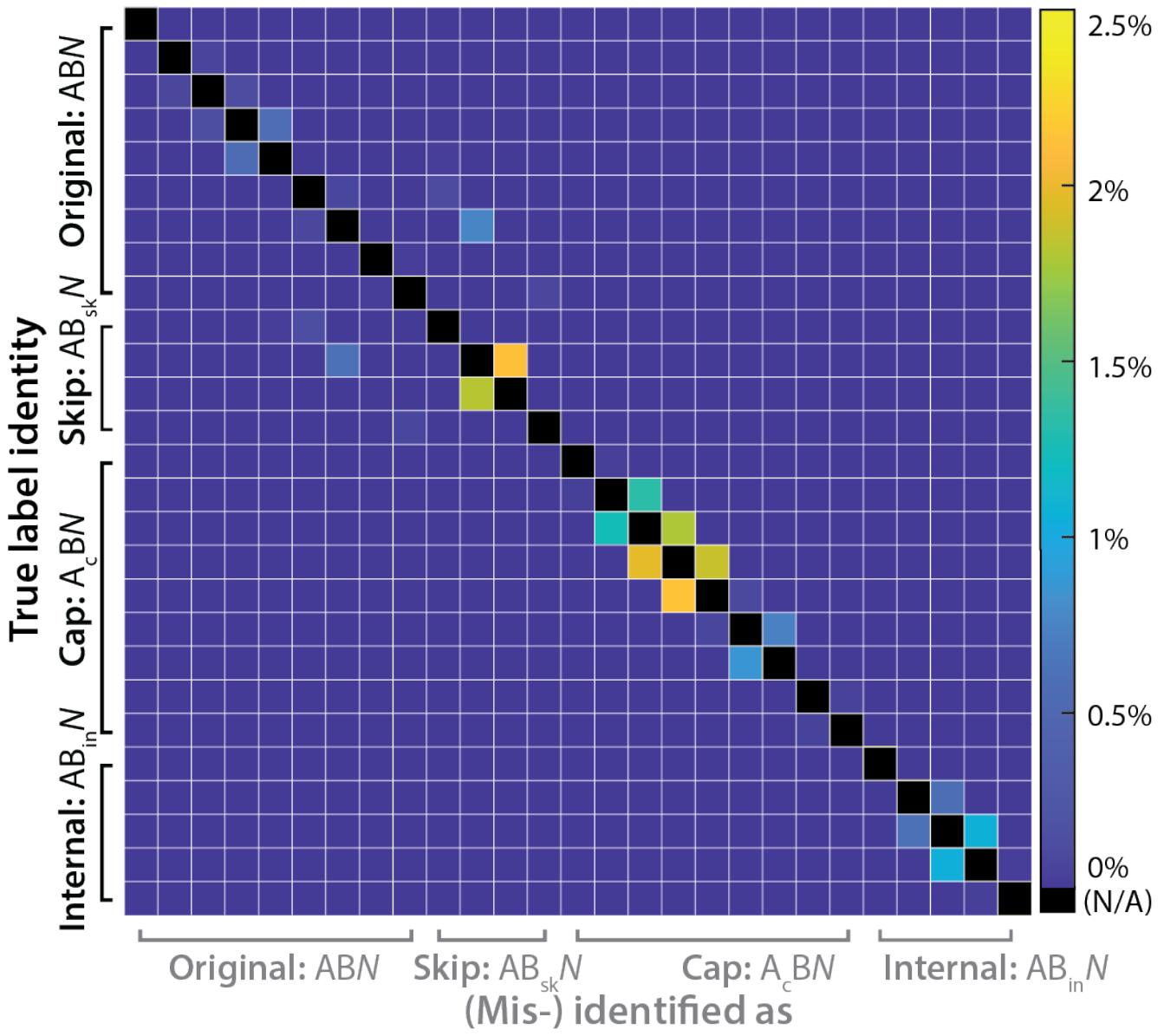

Figure S15. Statistical selection of a near-orthogonal FRETfluor set. One-tailed Gaussian overlap between each pair of clusters was calculated for the 27 selected constructs. The color bar cutoff is 2.5%.

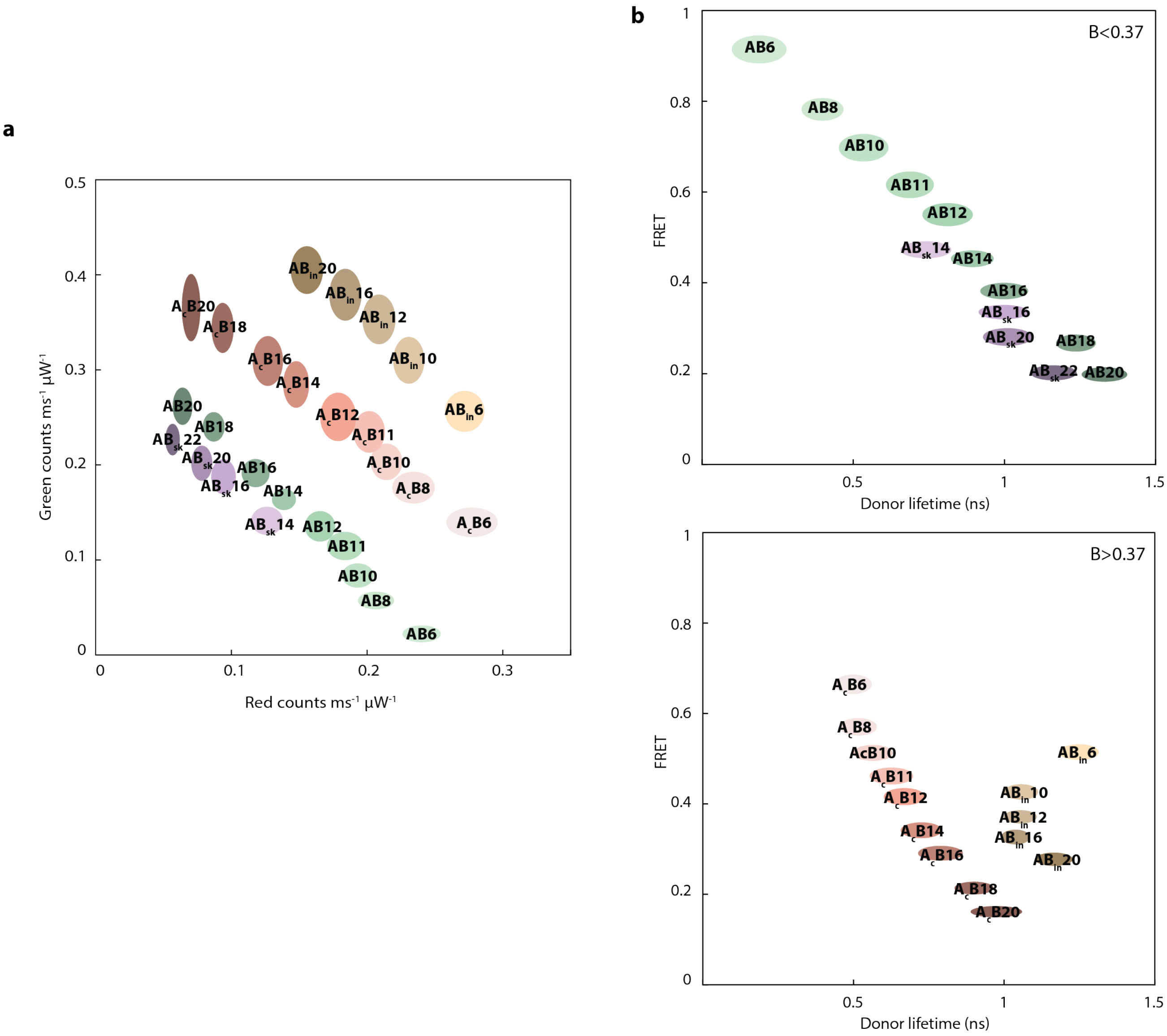

Figure S16. Cluster locations for 27-tag FRETfluor set. a) Red-green projection of data, plotted with the calculated center and 95% confidence interval oval. b) Two brightness slices of lifetime-FRET projection of data, plotted with the calculated center and 95% confidence interval oval.

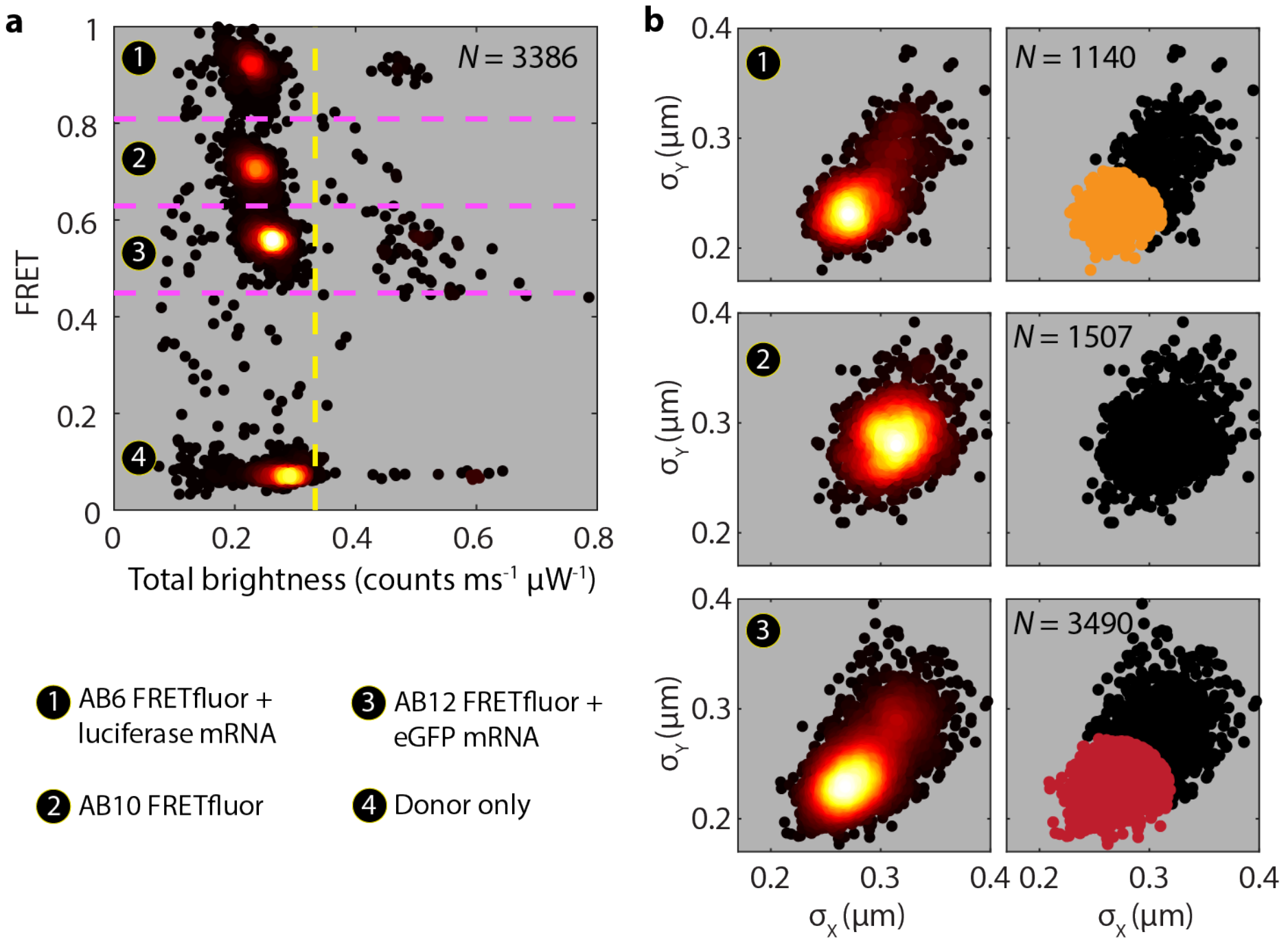

Figure S17. FRETfluor labeling and readout of an mRNA mixture. a) Total brightness of each observed level with minimum duration 100 ms; we excluded molecules that were too bright (above the yellow line; likely dimers) from our analysis. b) For each FRETfluor, scatter plots of the standard deviations of x and y positions during trapping are shown, calculated for 1000-photon bins. For the tags bound to a target, there are two populations corresponding to bound and unbound FRETfluors, which are not observed in the AB10 control. Hard clustering for the bound populations results in the points marked in orange and red, respectively (right); these correspond to the points shown in the main text Fig. 6b.

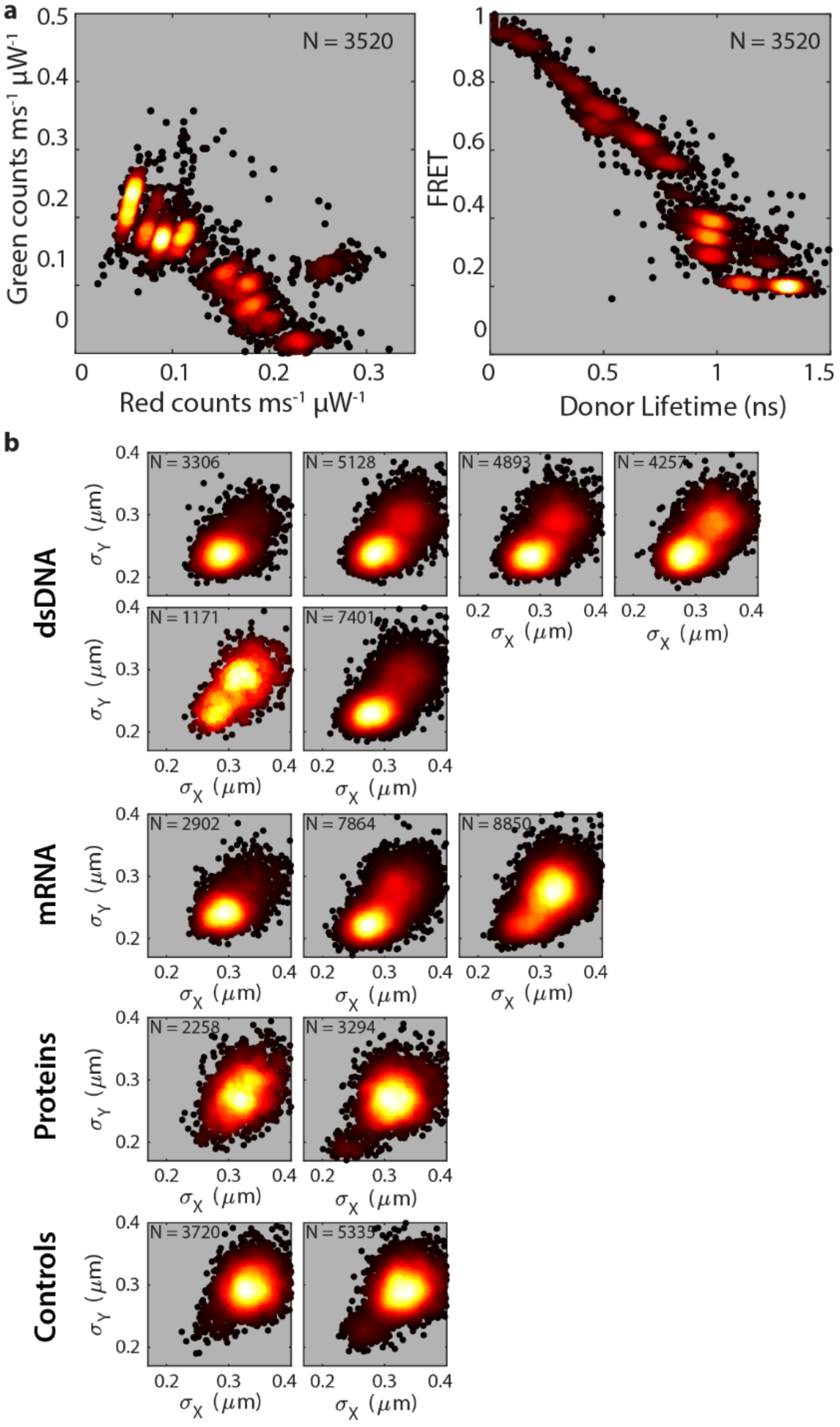

Figure S18. FRETfluor labeling and readout of a complex mixture of mRNA, dsDNA, and proteins. a) Red-green and lifetime-FRET projections of level-by-level data from FRETfluor-labeled dsDNA, mRNA, and proteins in a low-abundance mixture shows distinct clusters at the expected locations for each species. b) Scatter plot of standard deviation of position in *x* and *y* (calculated for 1000-photon bins) for trapped molecules of each species shows two populations in cases where the molecular weight change is significant (mRNA, dsDNA) but not for the controls. For the proteins, only one population is apparent, but it is trapped more tightly than the controls in each case. In all panels, points are colored according to the local relative scatter plot density.

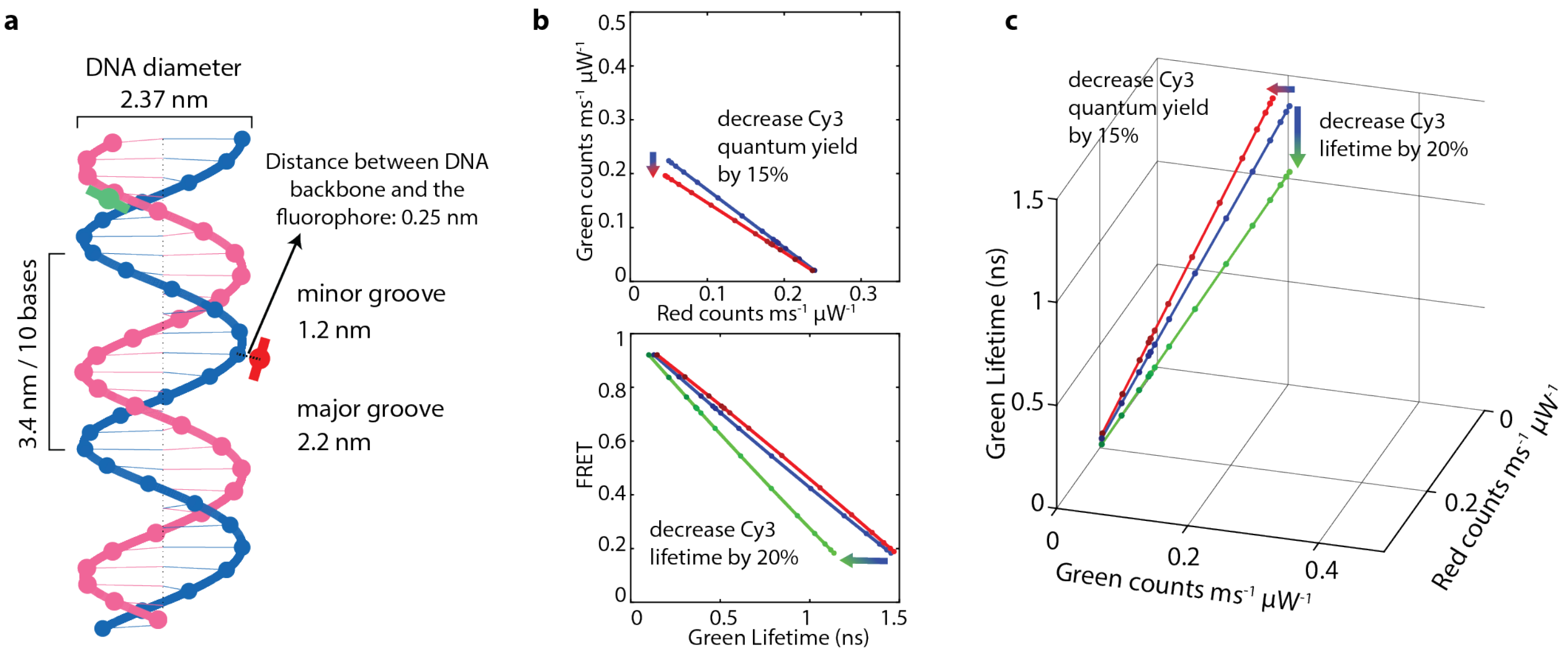

Figure S19. Simple geometrical model for FRET on DNA. a) Cartoon of a rigid DNA double helix with doubly-tethered Cy3 and Cy5 incorporated into the backbone (shown: spacing similar to AB7). b) The calculated spectroscopic signal when the Cy3 quantum yield is reduced by 15% (red) as compared to the original quantum yield (blue). The green line represents a Cy3 lifetime reduced by 20%. c) A three-dimensional view of the model results shows that decoupled changes to quantum yield and lifetime move the FRET curve in nearly orthogonal directions in the detection parameter space.

# Supplementary tables

## Supplementary Table S1: DNA oligomer sequences for FRETfluors and targeting bridges

| | | | |
|---|---|---|---|
| $A_{short}$ | 5' | GAT GAT GTC ATC GAC /iCy3/GCG CGA TAT TCC TAC TTA TGG CGG CTC TTC CCA G | 3' |
| A | 5' | GAT GAT GTC ATC GAC /iCy3/GCG CGA TAT TCC TAC TTA TGG CGG CTC TTC CCA GCG CTA ATC ACG TTC A | 3' |
| $A_c$ | 5' | /5Cy3/GAT GAT GTC ATC GAC /iCy3/GCG CGA TAT TCC TAC TTA TGG CGG CTC TTC CCA GCG CTA ATC ACG TTC A | 3' |
| B6 | 5' | CTG GGA AGA GCC GCC ATA AGT AGG AAT /iCy5/TCG CGC CGT CGA TGA CAT CAT C | 3' |
| B7 | 5' | CTG GGA AGA GCC GCC ATA AGT AGG AA/iCy5/A TCG CGC CGT CGA TGA CAT CAT C | 3' |
| B8 | 5' | CTG GGA AGA GCC GCC ATA AGT AGG A/iCy5/TA TCG CGC CGT CGA TGA CAT CAT C | 3' |
| B9 | 5' | CTG GGA AGA GCC GCC ATA AGT AGG /iCy5/ATA TCG CGC CGT CGA TGA CAT CAT C | 3' |
| B10 | 5' | CTG GGA AGA GCC GCC ATA AGT AG/iCy5/ AAT ATC GCG CCG TCG ATG ACA TCA TC | 3' |
| B11 | 5' | CTG GGA AGA GCC GCC ATA AGT A/iCy5/G AAT ATC GCG CCG TCG ATG ACA TCA TC | 3' |
| B12 | 5' | CTG GGA AGA GCC GCC ATA AGT /iCy5/GG AAT ATC GCG CCG TCG ATG ACA TCA TC | 3' |
| B13 | 5' | CTG GGA AGA GCC GCC ATA AG/iCy5/ AGG AAT ATC GCG CCG TCG ATG ACA TCA TC | 3' |
| B14 | 5' | CTG GGA AGA GCC GCC ATA A/iCy5/T AGG AAT ATC GCG CCG TCG ATG ACA TCA TC | 3' |
| B15 | 5' | CTG GGA AGA GCC GCC ATA /iCy5/GT AGG AAT ATC GCG CCG TCG ATG ACA TCA TC | 3' |
| B16 | 5' | CTG GGA AGA GCC GCC AT/iCy5/ AGT AGG AAT ATC GCG CCG TCG ATG ACA TCA TC | 3' |
| B17 | 5' | CTG GGA AGA GCC GCC A/iCy5/A AGT AGG AAT ATC GCG CCG TCG ATG ACA TCA TC | 3' |
| B18 | 5' | CTG GGA AGA GCC GCC /iCy5/TA AGT AGG AAT ATC GCG CCG TCG ATG ACA TCA TC | 3' |
| B19 | 5' | CTG GGA AGA GCC GC/iCy5/ ATA AGT AGG AAT ATC GCG CCG TCG ATG ACA TCA TC | 3' |
| B20 | 5' | CTG GGA AGA GCC G/iCy5/C ATA AGT AGG AAT ATC GCG CCG TCG ATG ACA TCA TC | 3' |
| $B_{sk}6$ | 5' | CTG GGA AGA GCC GCC ATA AGT AGG AAT A/iCy5/TC GCG CGT CGA TGA CAT CAT C | 3' |
| $B_{sk}10$ | 5' | CTG GGA AGA GCC GCC ATA AGT AGG /iCy5/AAT ATC GCG CGT CGA TGA CAT CAT C | 3' |
| $B_{sk}12$ | 5' | CTG GGA AGA GCC GCC ATA AGT A/iCy5/GG AAT ATC GCG CGT CGA TGA CAT CAT C | 3' |
| $B_{sk}13$ | 5' | CTG GGA AGA GCC GCC ATA AGT /iCy5/AGG AAT ATC GCG CGT CGA TGA CAT CAT C | 3' |
| $B_{sk}14$ | 5' | CTG GGA AGA GCC GCC ATA AG/iCy5/T AGG AAT ATC GCG CGT CGA TGA CAT CAT C | 3' |
| $B_{sk}16$ | 5' | CTG GGA AGA GCC GCC ATA/iCy5/ AGT AGG AAT ATC GCG CGT CGA TGA CAT CAT C | 3' |
| $B_{sk}20$ | 5' | CTG GGA AGA GCC GC/iCy5/C ATA AGT AGG AAT ATC GCG CGT CGA TGA CAT CAT C | 3' |
| $B_{sk}22$ | 5' | CTG GGA AGA GCC /iCy5/GCC ATA AGT AGG AAT ATC GCG CGT CGA TGA CAT CAT C | 3' |
| $B_{in}$ | 5' | TGA ACG TGA TTA GCG /3Cy3/ | 3' |
| $BR_{target}$ | 5' | AAC TGC CTG GTG ATA TGA ACG TGA TTA GCG | 3' |
| $BR_{off\text{-}target}$ | 5' | ATT CCT AAG TCT GAA TGA ACG TGA TTA GCG | 3' |
| Target(ssDNA) | 5' | TAT CAC CAG GCA GTT GAC AGT GTA GCA AGC TGT AAT AGA TGC GAG GGT CCA ATA C | 3' |
| $BR_{EGFP}$ | 5' | TGT TCT GCT GGT AGT GGT GAA CGT GAT TAG CG | 3' |
| $BR_{FLuc}$ | 5’ | CTC GGC GTA GGT GAT GTC TGA ACG TGA TTA GCG | 3’ |
| $BR_{OVA}$ | 5’ | TTG TTG ATC TGG GTT GAA CGT GAT TAG CG | 3’ |
| $BR_{off\text{-}mRNA}$ | 5' | AAC ACA TAA ATA AAA **TGA ACG TGA TTA GCG** | 3' |
| $BR_{amine}$ | 5' | /5AmMC6/**TGA ACG TGA TTA GCG** | 3' |
| $BR_{FBA1}$ | 5' | <u>CAC CCT TGA TGG AAG CAT TTT GAC CTA</u> AA**T GAA CGT GAT TAG CG** | 3' |
| $BR_{CDC19}$ | 5' | <u>CAA CAT CGT TGG TGG TGG TAC CAG T</u>AA A**TG AAC GTG ATT AGC G** | 3' |
| $BR_{ENO2}$ | 5' | <u>AGC AGC AGC GGC TCT AGC AGC GGC C</u>AA A**TG AAC GTG ATT AGC G** | 3' |
| $BR_{TSA1}$ | 5' | <u>AAA GGC TAG GAC AAC GTA CTT ACC C</u>AA A**TG AAC GTG ATT AGC G** | 3' |
| $BR_{RPL5}$ | 5' | <u>CTT CTC TTC TTC TTC TGA AAG GAG T</u>AA A**TG AAC GTG ATT AGC G** | 3' |
| $BR_{SSA3}$ | 5' | <u>CCA ACC TAT TAT CAA AAT CTT CAC C</u>AA A**TG AAC GTG ATT AGC G** | 3' |

**KEY for Supplementary Table S1:**

X: Base is opposite to Cy3 after annealing
<u>XX</u>: No base immediately opposite to Cy3, so the Cy3 will be more exposed to solvent
<u>X/iCy5/X</u>: No base immediately opposite to Cy5, so the Cy5 will be more exposed to solvent
**XXX**/**XXX**: Complementary sequences are shown in the same color
XXX: Complementary to the target mRNA, also highlighted in gray (see SI Note S4)
<u>XXX</u>: Complementary to the target dsDNA

## Supplementary Table S2: Lifetime fitting of donor-only constructs

| **Construct (population)** | **Brightness** (counts $\mu s^{-1}$ $\mu W^{-1}$) | **1-exp Lifetime** (ns) | **2-exp Lifetimes #1** (% population; | ns) | **2-exp Lifetimes #2** (% population; | ns) |
|---|---|---|---|---|---|---|
| **AB0** | 0.31 ± 0.012 | 1.60 ± 0.03 | 41% ± 4.0% | 0.69 ± 0.05 | 59% ± 4.0% | 1.80 ± 0.03 |
| **AB$_{sk}$0** | 0.26 ± 0.007 | 1.25 ± 0.03 | 37% ± 4.5% | 0.64 ± 0.07 | 63% ± 4.5% | 1.41 ± 0.05 |
| **AB$_{in}$0** (AB$_{in}$0) | 0.56 ± 0.015 | 1.51 ± 0.03 | 44% ± 1.1% | 0.65 ± 0.02 | 56% ± 1.1% | 1.76 ± 0.02 |
| (bridge Cy3 only) | 0.24 ± 0.010 | 1.40 ± 0.03 | 47% ± 4.2% | 0.55 ± 0.03 | 53% ± 4.2% | 1.65 ± 0.04 |
| **A$_c$B0** (A$_c$B0) | 0.40 ± 0.011 | 1.07 ± 0.03 | 59% ± 1.4% | 0.50 ± 0.03 | 41% ± 1.4% | 1.36 ± 0.02 |
| (cap Cy3 only) | 0.17 ± 0.007 | 0.52 ± 0.02 | 70% ± 11% | 0.32 ± 0.06 | 30% ± 11% | 0.72 ± 0.08 |
| (Q-A$_c$B0) | 0.32 ± 0.013 | 0.82 ± 0.03 | 62% ± 5% | 0.45 ± 0.04 | 38% ± 5% | 1.07 ± 0.04 |

## Supplementary Table S3: FRETfluor cluster locations

| Tag | FRET unitless, [0,1] | Donor Lifetime (ns) | Red Brightness (counts $\mu s^{-1}$ $\mu W^{-1}$) | Green Brightness (counts $\mu s^{-1}$ $\mu W^{-1}$) |
|---|---|---|---|---|
| **AB6** | 0.916 ± 0.016 | 0.187 ± 0.045 | 0.240 ± 0.007 | 0.022 ± 0.004 |
| AB7 | 0.904 ± 0.016 | 0.270 ± 0.049 | 0.252 ± 0.010 | 0.027 ± 0.005 |
| **AB8** | 0.783 ± 0.013 | 0.398 ± 0.035 | 0.207 ± 0.007 | 0.057 ± 0.005 |
| AB9 | 0.743 ± 0.012 | 0.486 ± 0.050 | 0.210 ± 0.008 | 0.073 ± 0.006 |
| **AB10** | 0.698 ± 0.015 | 0.534 ± 0.041 | 0.193 ± 0.006 | 0.084 ± 0.006 |
| **AB11** | 0.616 ± 0.015 | 0.687 ± 0.040 | 0.184 ± 0.007 | 0.115 ± 0.007 |
| **AB12** | 0.550 ± 0.013 | 0.813 ± 0.042 | 0.165 ± 0.006 | 0.135 ± 0.008 |
| AB13 | 0.488 ± 0.011 | 0.797 ± 0.035 | 0.136 ± 0.006 | 0.143 ± 0.009 |
| **AB14** | 0.453 ± 0.009 | 0.895 ± 0.035 | 0.138 ± 0.004 | 0.167 ± 0.006 |
| AB15 | 0.482 ± 0.017 | 0.839 ± 0.042 | 0.148 ± 0.007 | 0.159 ± 0.010 |
| **AB16** | 0.381 ± 0.009 | 1.001 ± 0.040 | 0.118 ± 0.005 | 0.191 ± 0.007 |
| AB17 | 0.342 ± 0.012 | 1.034 ± 0.038 | 0.107 ± 0.006 | 0.207 ± 0.009 |
| **AB18** | 0.266 ± 0.009 | 1.238 ± 0.033 | 0.087 ± 0.004 | 0.240 ± 0.008 |
| AB19 | 0.214 ± 0.007 | 1.234 ± 0.030 | 0.067 ± 0.004 | 0.244 ± 0.013 |
| **AB20** | 0.197 ± 0.008 | 1.333 ± 0.038 | 0.064 ± 0.004 | 0.262 ± 0.010 |
| $AB_{sk}6$ | 0.962 ± 0.012 | 0.125 ± 0.047 | 0.272 ± 0.010 | 0.011 ± 0.003 |
| $AB_{sk}10$ | 0.776 ± 0.014 | 0.417 ± 0.033 | 0.210 ± 0.008 | 0.061 ± 0.004 |
| $AB_{sk}12$ | 0.551 ± 0.008 | 0.676 ± 0.023 | 0.149 ± 0.003 | 0.121 ± 0.004 |
| $AB_{sk}13$ | 0.480 ± 0.009 | 0.699 ± 0.025 | 0.123 ± 0.004 | 0.133 ± 0.004 |
| **$AB_{sk}14$** | 0.472 ± 0.009 | 0.741 ± 0.042 | 0.126 ± 0.006 | 0.141 ± 0.007 |
| **$AB_{sk}16$** | 0.334 ± 0.008 | 0.998 ± 0.042 | 0.094 ± 0.005 | 0.188 ± 0.009 |
| **$AB_{sk}20$** | 0.279 ± 0.010 | 1.011 ± 0.041 | 0.078 ± 0.004 | 0.202 ± 0.009 |
| **$AB_{sk}22$** | 0.200 ± 0.008 | 1.164 ± 0.038 | 0.057 ± 0.003 | 0.227 ± 0.008 |
| **$A_cB6$** | 0.665 ± 0.011 | 0.498 ± 0.031 | 0.277 ± 0.009 | 0.140 ± 0.008 |
| **$A_cB8$** | 0.571 ± 0.010 | 0.515 ± 0.031 | 0.234 ± 0.008 | 0.176 ± 0.008 |
| **$A_cB10$** | 0.513 ± 0.009 | 0.564 ± 0.033 | 0.214 ± 0.006 | 0.203 ± 0.010 |
| **$A_cB11$** | 0.462 ± 0.009 | 0.626 ± 0.036 | 0.202 ± 0.006 | 0.235 ± 0.011 |
| **$A_cB12$** | 0.416 ± 0.009 | 0.669 ± 0.034 | 0.179 ± 0.007 | 0.251 ± 0.013 |
| **$A_cB14$** | 0.342 ± 0.009 | 0.725 ± 0.034 | 0.148 ± 0.005 | 0.285 ± 0.012 |
| **$A_cB16$** | 0.291 ± 0.008 | 0.789 ± 0.037 | 0.127 ± 0.006 | 0.309 ± 0.013 |
| **$A_cB18$** | 0.214 ± 0.007 | 0.900 ± 0.032 | 0.094 ± 0.004 | 0.344 ± 0.013 |
| **$A_cB20$** | 0.161 ± 0.006 | 0.975 ± 0.043 | 0.070 ± 0.003 | 0.365 ± 0.018 |
| **$AB_{in}6$** | 0.515 ± 0.009 | 1.249 ± 0.033 | 0.272 ± 0.007 | 0.256 ± 0.011 |
| $AB_{in}8$ | 0.457 ± 0.006 | 1.132 ± 0.019 | 0.239 ± 0.005 | 0.284 ± 0.007 |
| **$AB_{in}10$** | 0.427 ± 0.009 | 1.057 ± 0.028 | 0.231 ± 0.006 | 0.310 ± 0.012 |
| $AB_{in}11$ | 0.403 ± 0.005 | 1.047 ± 0.018 | 0.220 ± 0.004 | 0.327 ± 0.007 |
| **$AB_{in}12$** | 0.372 ± 0.007 | 1.054 ± 0.027 | 0.209 ± 0.006 | 0.353 ± 0.013 |
| $AB_{in}14$ | 0.344 ± 0.005 | 1.017 ± 0.019 | 0.203 ± 0.004 | 0.387 ± 0.008 |
| **$AB_{in}16$** | 0.327 ± 0.008 | 1.038 ± 0.024 | 0.184 ± 0.006 | 0.379 ± 0.014 |

| | | | | | | | | | | | | |
|---|---|---|---|---|---|---|---|---|---|---|---|---|
| $AB_{in}18$ | 0.290 | ± | 0.005 | 1.088 | ± | 0.017 | 0.159 | ± | 0.004 | 0.388 | ± | 0.008 |
| **$AB_{in}20$** | 0.278 | ± | 0.007 | 1.166 | ± | 0.031 | 0.156 | ± | 0.006 | 0.405 | ± | 0.013 |

## Supplementary Table S4: Heterogeneous mixture details

| Target | Target type | FRETfluor label | Bridge name | Notes |
|---|---|---|---|---|
| FBA1 | dsDNA | AB6 | $BR_{FBA1}$ | RT-PCR product for fructose 1,6-bisphosphate aldolase |
| EGFP | mRNA | AB8 | $BR_{EGFP}$ | mRNA for enhanced Green Fluorescent Protein |
| CDC19 | dsDNA | AB10 | $BR_{CDC19}$ | RT-PCR product for pyruvate kinase |
| ENO2 | dsDNA | AB11 | $BR_{ENO2}$ | RT-PCR product for phosphopyruvate hydratase |
| TSA1 | dsDNA | AB12 | $BR_{TSA1}$ | RT-PCR product for thioredoxin peroxidase |
| RPL5 | dsDNA | AB14 | $BR_{RPL5}$ | RT-PCR product for a ribosomal 60S subunit |
| FLuc | mRNA | AB16 | $BR_{FLuc}$ | mRNA for firefly luciferase protein |
| Ydj1 | protein | AB18 | $BR_{amine}$ | Cytosolic class A J-domain protein |
| OVA | mRNA | AB20 | $BR_{OVA}$ | mRNA for ovalbumin |
| Pab1 | protein | $A_cB6$ | $BR_{amine}$ | Poly(A) binding protein from yeast |
| SSA3 | dsDNA | $AB_{sk}16$ | $BR_{SSA3}$ | RT-PCR product for heat shock protein Hsp70 |
| control | - | $AB_{sk}20$ | None | FRETfluor only |
| control | - | $AB_{sk}22$ | None | FRETfluor only |

## Supplementary Table S5: Target genes and RT-PCR primers

| Gene | Forward Primer (5'-3') | Reverse Primer (5'-3') | Reference Sequence (S288C, coding DNA)[1,2] |
|---|---|---|---|
| FBA1 | TCGCTGGTAAGGGTATCTCTAA | CCGTGGAAGACCAAGAACAA | https://www.yeastgenome.org/locus/S000001543/sequence |
| CDC19 | ACACCAAGGGTCCAGAAATC | TCACCTCTGGCAACCATAAC | https://www.yeastgenome.org/locus/S000000036/sequence |
| ENO2 | CTAACGCTATCTTGGGTGTCTC | GGAGTGGTACATGTCAGCTAAT | https://www.yeastgenome.org/locus/S000001217/sequence |
| TSA1 | GACGAAGTCTCCTTGGACAAATA | GGCAGCTTCGAAGTATTCCT | https://www.yeastgenome.org/locus/S000004490/sequence |
| RPL5 | CTCTGCTTACTCCTCTCGTTTC | CTGGGTCAATTTCTTCGGTTTC | https://www.yeastgenome.org/locus/S000006052/sequence |
| SSA3 | CCGCAGGAGACACTCATTTA | CCTTCGAGAGCCTACCTTTATC | https://www.yeastgenome.org/locus/S000000171/sequence |

# Supplementary Notes

## Supplementary Note S1: Sequence- and attachment-dependent photophysics of Cy3

Physical and chemical properties of the nano-environment of a fluorophore, and in particular the resulting dielectric environment, strongly influence fluorophore photophysics. Different solvents, substrates, and attachment chemistries therefore can alter observed photophysical properties from their baseline values; conversely, small modifications to the local environment can be used to intentionally alter these properties, as we have done in this work. Traditionally, the brightness or quantum yield $\Phi$ of a fluorophore is expected to change proportionally with the observed lifetime $\tau$, since

$$\tau = \frac{1}{k_r + k_{nr}} \tag{1}$$

and

$$\Phi = \frac{k_r}{k_r + k_{nr}} \tag{2}$$

and the native radiative rate $k_r$ of a fluorophore is assumed to be constant while only $k_{nr}$ changes. However, if the dielectric environment of the fluorophore chances due to changing exposure to solvent or to changes in the chemical surroundings, $k_r$ can also change, decoupling $\tau$ and $\Phi$.[1]

We measured the brightness and fluorescence lifetime of Cy3 alone for each type of construct used in this work. In all naming conventions used here, a “0” denotes the lack of a Cy5 dye on the B strand. Supplementary Figure S1 shows aggregated single-molecule data for Cy3-only complexes. Constructs with two Cy3s show a minimum of three populations (one each for the Cy3s, one for the combined signal). We observe that the skipped constructs $AB_{sk}0$ show slightly reduced lifetime and brightness, possibly due to additional conformational flexibility and solvent exposure. The additional Cy3 on the bridge is also dimmer with a shorter lifetime, but clearly distinct from the $AB_{sk}0$ construct. The total $AB_{in}0$ signal is a near-perfect sum of the bridge Cy3 and the original AB0.

The $A_cB0$ construct is more complicated, likely due to base stacking or other conformational changes induced by the Cy3 at the 5’ end of the A strand. A very dim population appears to correspond to the cap Cy3 only, as evidenced by transitions to and from the total $A_cB0$ signal population. The additional population is labeled with a “Q” because it appears to be a slightly quenched version of the $A_cB0$ population, although the reason for this population is not understood. We do not see FRETfluor signals for the $A_cBN$ series that are consistent with the Q population acting as a donor; it is rarely seen.

All donor-only population means and standard deviations for brightness and 1-exp lifetime fits, along with the 2-exp lifetime fits for comparison, are provided in Supplementary Table S2. For the 2-exp lifetimes, photons from all levels in the 1-exp populations shown in the scatter plots were aggregated into a single decay for each data set, for which a 2-exp lifetime fit was performed. Lifetimes, standard deviations, and fractional populations are given for a set of fits across a minimum of seven data sets each (max 15); outliers would have been excluded but the 2-exp results were consistent and did not contain outliers.

Critically, we observe that $\tau$ and $\Phi$ indeed appear to vary independently for most Cy3-only populations. This is useful in the context of this work, as explained in Supplementary Note S5 and Supplementary Fig. S13, because this allows the FRET curve to be moved in different directions.

## Supplementary Note S2: Raw trapping events

Supplementary Fig. S2 shows raw trapping events from a mixture including 27 FRETfluor labels. As described in *Methods*, change-point detection is used to identify all levels in the data on the basis of maximum likelihood changes in brightness. After clustering of the levels, the main populations for each of the labels can be identified, and those cluster locations and variance (SI Table S3) are used to classify all subsequent data. Events are automatically classified based on the type(s) of levels passing a duration filter (>150 ms) that they contain:

Most events contain a single state, but some also contain “unidentified” levels that do not match any cluster. We found that many (if not most) unidentified levels within assigned events represent “allowed transition levels” for that FRETfluor species, even though we did not explicitly analyze these levels’ clusters. This is evident because the “allowed transition” states (a) sometimes show up multiple times within one event and (b) have combinations of measured parameters that are plausibly explained for that construct (for example, by blinking of a donor). Photophysical parameters within individual levels are generally stable over time. In cases where non-allowed transitions are observed without a background level indicating an empty trap between events, it is likely that one trapped molecule was randomly replaced with another during the event.

Here we highlight a few selected events to illustrate these effects:

For example, in the second row between 120-123 seconds, which shows an event from the tag AB8, the donor and acceptor are on during the event. Then, at 123 sec, the acceptor photobleaches, dropping the red brightness to nearly zero, and the green brightness level goes up. The new green-only level is consistent with the properties of the AB0 donor only construct as characterized in Supplementary Fig. S1.

Another example is the event in the first row at 15-25 seconds, which shows an event from the tag $AB_{in}10$, where both donors are on during the first part of the event. Then at 21 sec, the donor on the bridge blinks, leaving the red brightness the same and the green brightness lower than before. The spectroscopic information for the second part of this event is similar to an AB10 label. However, since there is no gap between these two states, we conclude that this blinking state also belongs to tag $AB_{in}10$.

In the second row at 62-70 seconds, the first part of the event shows both donors on $A_cB10$ are on, while the second part (at 65 sec) shows that the donor on the cap is off. This leaves the green brightness lower while the red brightness does not change much. The spectroscopic information for the second part of this event is also similar AB10. However, since the gap between these two states is extremely short, we also assign this level to tag $A_cB10$.

There are also several events indicating that there is very weak energy transfer between the second donor and the acceptor on the main strand. In the second row between 10-22 seconds, the first part of the event shows both donors and the acceptor for $AB_{in}20$ are all on while the second part (at 20 sec) shows that the donor on the bridge has blinked off. The photobleaching of the donor on the bridge leaves both the red and green brightness lower than before.

## Supplementary Note S3: Trapping event rate and limit of detection

To determine the lowest working concentrations at which FRETfluors could reasonably be detected, we characterized the trapping event rate in our ABEL trap setup. We acquired data for two separate dilution series of the FRETfluor AB12. We tested concentrations from 5 pM down to 5 fM (Exp 1: 50 fM, 1 pM, 5 pM; Exp 2: 5 pM, 500 fM, 100 fM, 50 fM, 25 fM). We collected three 5-minute long data sets for each sample and rinsed the sample cell thoroughly between uses with DI water (>5x sequential volume rinses each). In the first experiment (Supplementary Fig. S3a), samples were tested in order of increasing concentration. In the second experiment (Supplementary Fig. S3b), samples were tested in order of decreasing concentration. The consistent results obtained from these two experiments suggests that washing out the ABEL trap cell effectively removes the previous sample to below detectable concentrations.

In analyzing the data, trapping events were defined as consecutive 10-ms bins above background with a minimum total level duration of 150 ms and brightness and lifetime values within 3.5 standard deviations of the AB12 population (Supplementary Table S3). Zero events were recorded across buffer-only data sets. Linear fits with intercept = 0 were fit to each data set, with slopes ~0.1 events $s^{-1}$ $pM^{-1}$.

Assuming Poisson-distributed trapping events, an expected value of 3 events during the measurement window is required to reach 95% confidence of nonzero rate (5% probability of recording 0 counts):

$$P(0) = \frac{\langle \#\text{events} \rangle^0 e^{\langle \#\text{events} \rangle}}{\langle \#\text{events} \rangle !} = 0.05 \tag{3}$$

This constraint yields ⟨#events⟩=3 counts. Therefore, in a 15 min measurement window ($t$ = 900 s), the minimum statistically detectable concentration would be:

$$C = \frac{\langle \#events \rangle}{kt} = \frac{3 \text{ counts}}{(0.1 \,{}^{\text{counts}}/_{\text{s}\cdot\text{pM}})(900 \text{ s})} = 33 \text{ fM} \tag{4}$$

We routinely work at hundreds of fM; in the present work the 27-component FRETfluor mixture data was acquired at ~70 fM per component, and the 13-component mixed biomolecular sample was acquired at ~350 fM per component.

## Supplementary Note S4: Sequence of mRNAs and FRETfluor binding sites

The three mRNAs used in this work encode the enhanced Green Fluorescent Protein (EGFP; 996 nt), firefly luciferase (FLuc; 1929 nt), and ovalbumin (OVA; 1438 nt), which include proprietary tail sequences to help stabilize the mRNA against degradation. The sequences are given as follows:

**EGFP mRNA:**

```
[tail*]AUGGUGAGCAAGGGCGAGGAGCUGUUCACCGGGGUGGUGCCCAUCCUGGUCGAGCUGGACGGC
GACGUAAACGGCCACAAGUUCAGCGUGUCCGGCGAGGGCGAGGGCGAUGCCACCUACGGCAAGCUGACCC
UGAAGUUCAUCUGCACCACCGGCAAGCUGCCCGUGCCCUGGCCCACCCUCGUGACCACCCUGACCUACGG
CGUGCAGUGCUUCAGCCGCUACCCCGACCACAUGAAGCAGCACGACUUCUUCAAGUCCGCCAUGCCCGAA
GGCUACGUCCAGGAGCGCACCAUCUUCUUCAAGGACGACGGCAACUACAAGACCCGCGCCGAGGUGAAGU
```

UCGAGGGCGACACCCUGGUGAACCGCAUCGAGCUGAAGGGCAUCGACUUCAAGGAGGACGGCAACAUCCU
GGGGCACAAGCUGGAGUACAACUACAACAGCCACAACGUCUAUAUCAUGGCCGACAAGCAGAAGAACGGC
AUCAAGGUGAACUUCAAGAUCCGCCACAACAUCGAGGACGGCAGCGUGCAGCUCGCCGACCACUACCAGC
AGAACACCCCCAUCGGCGACGGCCCCGUGCUGCUGCCCGACAACCACUACCUGAGCACCCAGUCCGCCCU
GAGCAAAGACCCCAACGAGAAGCGCGAUCACAUGGUCCUGCUGGAGUUCGUGACCGCCGCCGGGAUCACU
CUCGGCAUGGACGAGCUGUACAAGUAA[tail*]

**FLuc mRNA:**

[tail*]AUGGAGGACGCCAAGAACAUCAAGAAGGGCCCCGCCCCCUUCUACCCCCUGGAGGACGGCACC
GCCGGCGAGCAGCUGCACAAGGCCAUGAAGCGGUACGCCCUGGUGCCCGGCACCAUCGCCUUCACCGACG
CCCACAUCGAGGUGGACAUCACCUACGCCGAGUACUUCGAGAUGAGCGUGCGGCUGGCCGAGGCCAUGAA
GCGGUACGGCCUGAACACCAACCACCGGAUCGUGGUGUGCAGCGAGAACAGCCUGCAGUUCUUCAUGCCC
GUGCUGGGCGCCCUGUUCAUCGGCGUGGCCGUGGCCCCCGCCAACGACAUCUACAACGAGCGGGAGCUGC
UGAACAGCAUGGGCAUCAGCCAGCCCACCGUGGUGUUCGUGAGCAAGAAGGGCCUGCAGAAGAUCCUGAA
CGUGCAGAAGAAGCUGCCCAUCAUCCAGAAGAUCAUCAUCAUGGACAGCAAGACCGACUACCAGGGCUUC
CAGAGCAUGUACACCUUCGUGACCAGCCACCUGCCCCCGGCUUCAACGAGUACGACUUCGUGCCCGAGA
GCUUCGACCGGGACAAGACCAUCGCCCUGAUCAUGAACAGCAGCGGCAGCACCGGCCUGCCCAAGGGCGU
GGCCCUGCCCCACCGGACCGCCUGCGUGCGGUUCAGCCACGCCCGGGACCCCAUCUUCGGCAACCAGAUC
AUCCCCGACACCGCCAUCCUGAGCGUGGUGCCCUUCCACCACGGCUUCGGCAUGUUCACCACCCUGGGCU
ACCUGAUCUGCGGCUUCCGGGUGGUGCUGAUGUACCGGUUCGAGGAGGAGCUGUUCCUGCGGAGCCUGCA
GGACUACAAGAUCCAGAGCGCCCUGCUGGUGCCCACCCUGUUCAGCUUCUUCGCCAAGAGCACCCUGAUC
GACAAGUACGACCUGAGCAACCUGCACGAGAUCGCCAGCGGCGGCGCCCCCCUGAGCAAGGAGGUGGGCG
AGGCCGUGGCCAAGCGGUUCCACCUGCCCGGCAUCCGGCAGGGCUACGGCCUGACCGAGACCACCAGCGC
CAUCCUGAUCACCCCCGAGGGCGACGACAAGCCCGGCGCCGUGGGCAAGGUGGUGCCCUUCUUCGAGGCC
AAGGUGGUGGACCUGGACACCGGCAAGACCCUGGGCGUGAACCAGCGGGGCGAGCUGUGCGUGCGGGGCC
CCAUGAUCAUGAGCGGCUACGUGAACAACCCCGAGGCCACCAACGCCCUGAUCGACAAGGACGGCUGGCU
GCACAGCGGCGACAUCGCCUACUGGGACGAGGACGAGCACUUCUUCAUCGUGGACCGGCUGAAGAGCCUG
AUCAAGUACAAGGGCUACCAGGUGGCCCCCGCCGAGCUGGAGAGCAUCCUGCUGCAGCACCCCAACAUCU
UCGACGCCGGCGUGGCCGGCCUGCCCGACGACGACGCCGGCGAGCUGCCCGCCGCCGUGGUGGUGCUGGA
GCACGGCAAGACCAUGACCGAGAAGGAGAUCGUGGACUACGUGGCCAGCCAGGUGACCACCGCCAAGAAG
CUGCGGGGCGGCGUGGUGUUCGUGGACGAGGUGCCCAAGGGCCUGACCGGCAAGCUGGACGCCCGGAAGA
UCCGGGAGAUCCUGAUCAAGGCCAAGAAGGGCGGCAAGAUCGCCGUGUGA[tail*]

**OVA mRNA:**

[tail*]AUGGGCAGCAUCGGCGCCGCCAGCAUGGAGUUCUGCUUCGACGUGUUCAAGGAGCUGAAGGUG
CACCACGCCAACGAGAACAUCUUCUACUGCCCCAUCGCCAUCAUGAGCGCCCUGGCCAUGGUGUACCUGG
GCGCCAAGGACAGCACCCGGACCCAGAUCAACAAGGUGGUGCGGUUCGACAAGCUGCCCGGCUUCGGCGA
CAGCAUCGAGGCCCAGUGCGGCACCAGCGUGAACGUGCACAGCAGCCUGCGGGACAUCCUGAACCAGAUC
ACCAAGCCCAACGACGUGUACAGCUUCAGCCUGGCCAGCCGGCUGUACGCCGAGGAGCGGUACCCCAUCC
UGCCCGAGUACCUGCAGUGCGUGAAGGAGCUGUACCGGGGCGGCCUGGAGCCCAUCAACUUCCAGACCGC
CGCCGACCAGGCCCGGGAGCUGAUCAACAGCUGGGUGGAGAGCCAGACCAACGGCAUCAUCCGGAACGUG
CUGCAGCCCAGCAGCGUGGACAGCCAGACCGCCAUGGUGCUGGUGAACGCCAUCGUGUUCAAGGGCCUGU
GGGAGAAGACCUUCAAGGACGAGGACACCCAGGCCAUGCCCUUCCGGGUGACCGAGCAGGAGAGCAAGCC
CGUGCAGAUGAUGUACCAGAUCGGCCUGUUCCGGGUGGCCAGCAUGGCCAGCGAGAAGAUGAAGAUCCUG
GAGCUGCCCUUCGCCAGCGGCACCAUGAGCAUGCUGGUGCUGCUGCCCGACGAGGUGAGCGGCCUGGAGC
AGCUGGAGAGCAUCAUCAACUUCGAGAAGCUGACCGAGUGGACCAGCAGCAACGUGAUGGAGGAGCGGAA
GAUCAAGGUGUACCUGCCCCGGAUGAAGAUGGAGGAGAAGUACAACCUGACCAGCGUGCUGAUGGCCAUG

```
GGCAUCACCGACGUGUUCAGCAGCAGCGCCAACCUGAGCGGCAUCAGCAGCGCCGAGAGCCUGAAGAUCA
GCCAGGCCGUGCACGCCGCCCACGCCGAGAUCAACGAGGCCGGCCGGGAGGUGGUGGGCAGCGCCGAGGC
CGGCGUGGACGCCGCCAGCGUGAGCGAGGAGUUCCGGGCCGACCACCCCUUCCUGUUCUGCAUCAAGCAC
AUCGCCACCAACGCCGUGCUGUUCUUCGGCCGGUGCGUGAGCCCCUGA[tail*]
```

```
    *Proprietary tail sequences (TriLink BioTechnologies)
```

The secondary structures of the mRNAs were predicted by RNAfold,[3,4] and the EGFP structure is shown in Supplementary Fig. S4 as an example. In each case, the target sequence is located in a high-confidence loop of the structure, highlighted here in gray and indicated for EGFP on the structure shown in Supplementary Fig. S4 by the location of the FRETfluor binding. All bridge sequences are given in Supplementary Table S1, with gray highlights indicating the complementary regions to the target mRNA sequences.

## Supplementary Note S5: Effects of salt and pH on FRETfluor properties

To investigate the sensitivity of FRETfluors to the surrounding solution environment, we measured the effects of varying pH and salt conditions on the measured spectroscopic parameters of a representative set of 10 FRETfluors. The FRETfluors were chosen to span the whole FRET range and included multiple members of each type of construct family.

We performed four different experiments using the same mixture of FRETfluors in different buffer conditions. For three experiments, the same buffer at three different pH (6.51, 7.4, 8.63) was used for the measurement (no salt added). For one experiment, we added 150 mM of NaCl to the neutral pH buffer to test the effects of salt.

To clearly visualize all the populations in the FRET versus donor lifetime projection, we first found the mean and the standard deviations of each tag in the data for all four experiments. In Supplementary Fig. S12, each tag's cluster location and spread are represented by an ellipse centered at its respective mean FRET and lifetime. The radius of the ellipse is twice the calculated standard deviation in each direction.

Our results show that there is no notable change in the FRETfluors signals within the pH range tested. However, at the higher salt concentration, the signal of the $AB_{in}6$ and $AB_{in}12$ is slightly shifted to lower donor lifetime. These tags contain an extra Cy3 outside of FRET range that is singly-tethered to the bridge strand, and therefore might be more influenced by the dielectric properties of the solution than the doubly-tethered Cy dyes. At the high salt concentration tested, many FRETfluors were also slightly dimmer. At high salt, the original, cap, and internal modified tags were each about 8-10% dimmer. The skip modified tags were only 2-4% dimmer. This further suggests that salt might be differentially influencing the photophysics of constructs, indicating that FRETfluor clusters should be carefully calibrated across all environmental conditions to be used, and/or that sets of FRETfluors that are robustly mutually identifiable even in the presence of such shifts should be selected for use.

## Supplementary Note S6: Identification of FRETfluors depends on # photons available

For this analysis, we parsed the data from each level into successive groups of *M* photons each (discarding remainder), with each group of *M* photons shown as one point on a scatter plot. We then measured the change in the standard deviation of the clusters for red brightness, green brightness, and green lifetime, as a function of the number of photons per point and calculated the effect this would have on misidentification of a typical FRETfluor, AB11, as compared to one or two close neighbors (AB10 and AB12) or as compared to more distal FRETfluor clusters (AB8, AB16).

## Supplementary Note S7: Simulation of energy transfer between Cy3 and Cy5 on dsDNA

To model the expected magnitude of changes to different parameters in the spectroscopic output of FRETfluors due to changes in donor or acceptor photophysical properties, we created a simple model of FRET for doubly-tethered Cy3 and Cy5 on double-stranded DNA (dsDNA) after published works.[5–7]

Supplementary Fig. S13 shows the simple geometrical model of a rigid DNA double helix with doubly-tethered Cy3 and Cy5 incorporated into the backbone. B-form DNA parameters were taken from Ref 7.[8] The diameter of the DNA is 2.37 nm, and the gap between base pairs is 0.34 nm (10 bp/turn). The distance between the DNA backbone and the fluorophore was set at 0.25 nm. The angular offset between the 3'-5' and the 5'-3' strand (-127°) is set by the minor groove height (1.2 nm) and helicity.

For simplicity, we assumed that both dyes could rotate freely ($\kappa^2 = 2/3$).[5] While some constriction of the dipole cone angles in FRETfluors is anticipated due to the backbone attachment of the Cy dyes, here our observed FRET curve for sequential constructs (for example, the AB*N* series shown in Fig. 4A) does not exhibit significant helix-dependent orientation effects. This result suggests that the dyes have relatively mobile orientation, so while kappa^2 may not be exactly 2/3, it does not vary substantially across constructs and is therefore likely to be close to 2/3.

To create different constructs that were representative of different dye spacings in our FRETfluors, the Cy3 (donor) was kept at the same location for each construct and the Cy5 (acceptor) was moved to different locations along the opposite strand.

Dye photophysical parameters were assigned according to measured values when possible. For example, we measured the single-exponential fit to the lifetime decay for Cy3 doubly-tethered to the DNA backbone at the single molecule level as 1.6 ns, as shown in Supplementary Fig. S1. Other parameters: Cy5 lifetime (1.7 ns), quantum yields $\Phi_{Cy3}$= 0.15 and $\Phi_{Cy5}$= 0.27.[9,10]

Briefly, we simulated a coupled energy transfer model where the time evolution of the probability of exciton residence on each fluorophore following absorption, $p = [p_{Cy3}(t), p_{Cy5}(t)]$ by one or the other fluorophore, $dp(t)/dt$, may be described by the master equation:

$$\frac{d\boldsymbol{p}(t)}{dt} = \boldsymbol{M}\boldsymbol{p}(t) \tag{5}$$

**M** is an excitation transition matrix with off-diagonal elements representing the pairwise energy transfer rates, $k_{ji}$, (row *i*, column *j*) between donor *j* and acceptor *i* pigments. Entries on the diagonal indicate the total rate of energy loss at that fluorophore, including both energy transfer to the other pigments and the rate of fluorescent emission at that site, $k_{ii}^{10}$:

$$M_{ii} = -\left(k_{ii}^{10} + \sum\nolimits_{i,j\neq i} k_{ji}\right) \tag{6}$$

To calculate initial values for the energy transfer rates, we used the Förster energy transfer equation[5,11] using absorption and emission spectra for Cy3 and Cy5:

$$k_{DA} = \frac{k_D^{10}\Phi_D\kappa^2}{R_{DA}^6}\left(\frac{9000\ln 10}{128\pi^5 N_A n^4}\right)\frac{\int Em_D\varepsilon_A\lambda^2 d\lambda}{\int Em_D\lambda^{-2}d\lambda} \tag{7}$$

For each initial condition *p*(0) = [0, 1] and *p*(0) = [1, 0], we can find the total probability of fluorescence emission from site *i* given excitation at site *j* by integrating over a long time (here: ~10 ns):

$$P_i^j \equiv \int_0^{+\infty} p_i^j(t)k_i^{10}\,\Phi_i dt \tag{8}$$

where $\Phi_i$ is the quantum yield of pigment *i*. The total probability of emission from fluorophore *i* given excitation at either fluorophore *j*, $P_i^j$, is then given according to the relative probabilities of each initial condition (determined by the absorbance probability at the excitation wavelength, 532 nm, for each fluorophore)

$$P_i = \sum_j A_j P_i^j \tag{9}$$

where $A_j$ is the absorption probability at 532 nm for fluorophore *j*. The total probability of emission from fluorophore *i*, $P_i$, is directly proportional to the photon emission rate, or brightness $B_i$, from that fluorophore, by some unknown constant (*vide infra*). The predicted emission spectrum is a weighted sum of the individual fluorophore spectra, where the relative brightness of each pigment is used to weigh its contribution to the spectrum:

$$Em(\lambda) = \frac{\sum_i B_i Em_i(\lambda)}{\sum_i B_i} \tag{10}$$

This spectrum is separated into two spectral windows that encompass the red channel and the green channel, respectively. In order to allow our simulation results to be compared to measured data, we multiplied each brightness channel by the transmission profile of our emission filters (see *Methods*). In total, this translated to 71.43% throughput for the green channel, *Em*(green), and 46.18% throughput for the red channel, *Em*(red). Next, we scaled these unitless brightnesses to be comparable to our experimental brightness data for the AB*N* constructs using an approximate scaling factor (3 counts / 10 ms / µW). These became the simulation outputs for green and red channel brightnesses, "Green counts" and "Red counts" per Supplementary Fig. S13b (top panel).

As with our experimental data, FRET values (Supplementary Fig. S13b, bottom panel) were calculated according to an uncorrected ratio between the red channel and the sum over both channels:

$$FRET = \frac{\Sigma Em(red)}{\Sigma Em(red) + \Sigma Em(green)} \tag{11}$$

Finally, we predicted the observable single-exponential fluorescence lifetime of the donor by constructing the fluorescence decay of that state, $g_{donor}(t)$, and then fitting a single exponential decay function to find the apparent lifetime (see Supplementary Fig. S13b, bottom panel):

$$g_{donor}(t) = \sum_j k_i^{10} \Phi_i \left( \frac{\sum_j A_j p_i^j(t)}{\sum_j A_j} \right) \quad (12)$$

We calculated green and red brightness, green lifetime, and FRET values for complexes with spacings from 6 to 20 bp.

## Supplementary Note S8: Effect of changing donor photophysical properties on FRET

We used our FRET simulation as described in Supplementary Note S4 above to model the expected magnitude of changes in spectroscopic signals we might measure for certain changes in photophysical properties of the donor Cy3. We separately modeled changes to the lifetime, $\tau$, and quantum yield, $\Phi$, of the donor, since our measurements of donor properties of different constructs indicated that changes in these variables across different constructs were not perfectly coupled (see Supplementary Note S1).

Per Supplementary Note S4, we experimentally observed donor brightnesses as low as 50% (for $A_{cap\text{-}only}B0$) compared to the brightest donor construct (AB0). A subtler reduction of ~15% brightness was observed for $AB_{sk}0$ as compared to AB0. Here, we take changes in brightness to reflect changes in quantum yield. For lifetimes, we observed reductions by up to a factor of three (for $A_{cap\text{-}only}B0$ as compared to AB0), and at minimum a 20% reduction ($AB_{sk}0$ as compared to AB0).

We modeled these minimal changes in $\Phi$ (15% reduction, red) and $\tau$ (20% reduction, green) using the simulation described in Supplementary Note S4 above to determine the predicted changes in our measured signals as compared to the original values (blue). As expected, both changes are most evident when FRET is low, and produce corresponding reductions in the observed brightness and lifetime for the lowest FRET states of approximately 15% and 20% for quantum yield and green lifetime, respectively (Supplementary Fig. S13b). We therefore expected that if populations with lower FRET could be clustered tightly within a few % or less, populations from the original constructs could be easily differentiated from the constructs with the modified donor properties.

These results also illustrate that the effects of changing $\Phi$ and $\tau$ are nearly orthogonal for the two data projections used throughout this work: The top panel is a brightness-brightness projection for the two color channels, while the bottom panel is a FRET- $\tau_{Green}$ projection. We note that the reduced Cy3 lifetime is apparent as a shifted green line only in the bottom panel of Supplementary Fig. S13b, and overlaps almost perfectly with the original blue line in the top panel (not shown due to overlap). Similarly, the reduced Cy3 quantum yield creates an obviously shifted red line top panel, but overlaps almost perfectly with the original blue line in the bottom panel.

In the three-dimensional view shown in Supplementary Fig. S13c, the near-orthogonality of these motions in Gr-R-$\tau$ space is clear. As discussed in Supplementary Note S4, traditionally $\Phi$ and $\tau$ are expected to be perfectly correlated under the assumption that the native radiative rate of a fluorophore does not change.[12] Here, our simulations illustrate the advantage of a situation where $\Phi$ and $\tau$ are not perfectly coupled: independently changing donor properties can move the highly correlated FRET curve around the multidimensional parameter space, providing opportunities to create many distinct constructs.

The shifts in the simulated FRET curve at low FRET both (1) roughly correspond to the donor-only changes of 15% QY decrease and 20% lifetime decrease that were selected based on the measured differences between AB0 and $AB_{sk}0$ (see SI Table S2), and (2) roughly match our experimental observations, as seen by comparing the results for low-FRET AB series clusters (for example, AB20) to their $AB_{sk}$ counterparts ($AB_{sk}20$, which is ~23% dimmer in green brightness and ~13% shorter donor lifetime per SI Table S3). This correlation supports the choice of simulation parameters.

## Supplementary Note S9: Instrument correction parameters

**FRET correction parameters:** To obtain accurate FRET values that can be replicated on different optical setups, a few corrections need to be performed. These correction factors can be experimentally measured using Alternating Laser Excitation (ALEX).[13–15] The three main FRET corrections are the donor leakage $\alpha$, normalization of effective quantum yields and detection efficiencies $\gamma$, and direct acceptor excitation by the green laser $\delta$. Using these correction factors, the corrected FRET is given by equation 13 where $\mathrm{F_{A|D}}$ is the acceptor fluorescence upon donor excitation and similarly $\mathrm{F_{D|D}}$ is the donor fluorescence upon donor excitation and $\mathrm{F_{A|A}}$ is the acceptor fluorescence upon acceptor excitation.

$$E_{FRET} = \frac{F_{A|D} - \alpha F_{D|D} - \delta F_{A|A}}{\gamma F_{D|D} + F_{A|D} - \alpha F_{D|D} - \delta F_{A|A}} \tag{13}$$

In the main figures for this paper, we show FRET values without performing any corrections (except for background subtraction). In this work, we are using FRET to create unique photophysical signatures and not to characterize a system or estimate distances using FRET. We have performed separate trapping experiments using ALEX on the same samples (data not shown). Using ALEX, we trapped a mixture containing 9 FRETfluors from the original design and determined the correction factors needed to arrive at accurate FRET values. Note that these ALEX experiments were performed using only one red and one green detection channel while the experiments from current work were done using two green and two red channels for anisotropy measurements. Regardless, these correction factors along with the *g* factor for our setup can be used to determine accurate FRET values for comparison across different instruments.

The correction parameters measured for our setup are: $\alpha = 0.08$, $\delta = 0.05$, and $\gamma = 0.67$.

**Determination of the *g* factor:** The *g* factor accounts for the difference in the detection efficiency of the detectors at the two polarizations. In our experiments, photons are first split by color, using a dichroic into red and green channels. Then, both the red and green channels are split using a polarizing beam splitter into parallel and perpendicular channels. Thus, we will have two g factors: one for the two green polarization channels ($g_{green}$) and one for the two red polarization channels ($g_{red}$). To experimentally determine the g factor, we use a concentrated sample of malachite green. First, a measurement is taken to record intensities of all four detectors using vertically polarized excitation (measurement #1). Then, the measurement is repeated using horizontally polarized excitation (measurement #2).

For simplicity, we use $D_s$ and $D_p$ to denote the detector that will accept $s$ (vertical to the table) and $p$ (horizontal to the table) polarized light as defined by the polarizing beam splitter. G factor is always used to scale the $D_s$ detector. Vertically polarized excitation is parallel to the $D_p$ detector and horizontally polarized excitation is parallel to the $D_s$ detector. Assuming the sample polarization should be the same whether we are using the horizontally or vertically polarized excitation, we have the following

(14)

relation where the left-hand side of the equation refers to the polarization determined using measurement #1 and right-hand side refers to measurement #2.

$$\frac{D_p^1 - gD_s^1}{D_p^1 - gD_s^1} = \frac{gD_s^2 - D_p^2}{gD_s^2 + D_p^2}$$

Solving for g results in:

$$g = \sqrt{\frac{D_p^1 D_p^2}{D_s^1 D_s^2}} \tag{15}$$

Our measured g factors are: $g_{green} = 1.0481$ and $g_{red} = 1.2998$.

# Supplementary references